\documentclass[a4paper,fleqn,usenatbib,useAMS]{mnras}

\usepackage{graphicx}   
\usepackage{amsmath}    
\usepackage{amssymb}    
\usepackage{subfigure}
\usepackage{epstopdf}

\usepackage[T1]{fontenc}
\usepackage{ae,aecompl}

\usepackage{txfonts}
\usepackage{mathptmx}

\title[Tidal Love numbers and moment-Love relations of polytropic stars]{Tidal Love numbers and moment-Love relations of polytropic stars}
\author[Kenny L. S. Yip, P. T. Leung]{Kenny L. S. Yip \thanks{Email address: \href{mailto:yiplongsang@gmail.com}{yiplongsang@gmail.com}},
P. T. Leung \thanks{Email address:
\href{mailto:ptleungphy@gmail.com}{ptleungphy@gmail.com}}
\\
Department of Physics, the Chinese University of Hong Kong,
Shatin, N.T., Hong Kong SAR}

\date{Accepted XXX. Received YYY; in original form ZZZ}

\pubyear{2017}

\begin{document}
\label{firstpage}
\pagerange{\pageref{firstpage}--\pageref{lastpage}}
\maketitle

\begin{abstract}
The physical significance of tidal deformation in astronomical systems has long been known. The recently discovered universal I-Love-Q relations, which connect moment of inertia, quadrupole tidal Love number, and spin-induced quadrupole moment of compact stars, also underscore the special role of tidal deformation in gravitational wave astronomy. Motivated by the observation that such relations also prevail in Newtonian stars and crucially depend on the stiffness of a star, we consider the tidal Love numbers of Newtonian polytropic stars whose stiffness is characterised by a polytropic index $n$. We first perturbatively solve the Lane-Emden equation governing the profile of polytropic stars through the application of the scaled delta expansion method and then formulate perturbation series for the multipolar tidal Love number about the two exactly solvable cases with $n=0$ and $n=1$, respectively. Making use of these two series to form a two-point Pad\'{e} approximant, we find an approximate expression of the quadrupole tidal Love number, whose error is less than $2.5 \times 10^{-5}$ per cent (0.39 per cent) for $n\in[0,1]$ ($n\in[0,3]$). Similarly, we also determine the mass moments for polytropic stars accurately.  Based on these findings, we are able to show that the I-Love-Q relations are in general stationary about the incompressible limit irrespective of the equation of state (EOS) of a star. Moreover, for the I-Love-Q relations, there is a secondary stationary point near $n\approx0.4444$, thus showing the insensitivity to $n$ for $n\in[0,1]$. Our investigation clearly tracks the universality of the I-Love-Q relations from their validity for stiff stars such as neutron stars to their breakdown for soft stars.  
\end{abstract}




\section{Introduction}
The phenomenon of tidal deformation is ubiquitous as well as
important in many astronomical systems such as coalescing binary
stars \citep[see, e.g.,][and references therein]{Lai_tidal}, our
own solar system and other planetary systems \citep[see,
e.g.,][and references therein]{Beuthe_2015}. In order to quantify
and analyse the tidal effects on a star (or a planet) due to other
companion stars (or planets), the concept of tidal Love numbers
was introduced \citep{Love_1909}, which is a measure of the ratio
of the induced multipole moment to an external multipole field
applied on the star (or planet) under consideration in the linear
response regime.
The tidal Love numbers of a star (or a planet) could depend
sensitively on its physical characteristics   (e.g.,  mass, radius
and elasticity) and its equation of state (EOS). \citep[see,
e.g.,][and references therein]{Brooker_1955,Damour_Nagar_2009,
Lattimer_2010,Beuthe_2015}. As a result, some proposals have been
put forward to infer the physical characteristics and the EOS of
neutron stars (NSs) and quark stars (QSs) from their tidal Love
numbers \citep[see, e.g.,][]{Damour_Nagar_2009, Lattimer_2010}.

Although the study of tidal Love numbers was first started more
than a century ago \citep{Love_1909}, it has regained proper
attention in this new millennium for its special role in
gravitational wave astronomy, which has been  realized recently
\citep{GW_first_2016}. \citet{Flanagan:08p021502} pointed out that
 tidal Love numbers of NSs could be inferred from
the associated gravitational wave signals emitted in the inspiral
stage of neutron star-black hole (NS-BH) and neutron star-neutron
star (NS-NS) binaries. In addition, \citet{YagiYunes2013,
YagiYunes2013Science} discovered the universal I-Love-Q
relations among the moment of inertia, the quadrupole tidal Love number
(deformability) and the spin-induced quadrupole moment of NSs and
QSs are almost independent of the EOS of these stars. Such
relations are remarkable in the sense that a measurement of one of
these three quantities could provide estimates of the other two with high
precision, and it is claimed that such relations could facilitate the
analysis of gravitational wave signals detected from 
mergers of compact star binaries \citep[see][and references
therein]{Yagi:2016bkt}.

The physical origin of these universal relations has attracted a
lot of interest since their discovery. \citet{YagiYunes2013}
attempted to attribute these universal relations of compact stars
to the no-hair theorem, which states that black holes do not have
interior structure, in the black hole limit. On the other hand,
several studies \citep{Sham_2015,Samson_fmode_ILoveQ,
ILoveQ_3,ILoveQ_4} indicate that the I-Love-Q relations indeed
originate from the facts that (i) compact stars (including both
NSs and QSs) are stiff enough to be considered as incompressible
during tidal deformation, and (ii) the I-Love-Q relations are
stationary and remarkably flat around the incompressible limit. As
a matter of fact, the I-Love-Q relations still prevail in the
Newtonian limit and their accuracies were found to worsen towards
the high compactness regime \citep{Sham_2015,ILoveQ_3,ILoveQ_4}.

Motivated by the findings of \citet{Sham_2015,ILoveQ_3,ILoveQ_4},
we study analytically in the present paper how the I-Love-Q
relations depend on the stiffness of respective stars under
Newtonian gravity by considering polytropic stars whose stiffness
can be varied continuously and is characterised by the polytropic
index $n$. As the spin-induced quadrupole moment is completely
specified by the quadrupole tidal Love number and the moment of inertia for
Newtonian stars, we focus our attention on the I-Love relation as
well as its multipolar generalisation in the present paper.  

Our main objective here is to obtain accurate approximants for the
I-Love relation of stable polytropic stars as a function of the
polytropic index $n$. Firstly, we make use of the scaled delta
expansion method proposed recently by \citet{Kenny_SDEM} to
find the density profile of polytropic stars. Secondly, a
perturbative expansion for the tidal Love numbers in terms of
moments of the mass density distribution for spherical Newtonian
stars is derived.  When such an expansion is applied to polytropic
stars whose density profile can be found approximately from the scaled delta
expansion method,
we succeed in finding perturbative corrections to the tidal Love
numbers of degree
 $l=2$ and $l=3$, up to the third order
about the incompressible limit $n=0$, and up to the first order
about the linear model $n=1$. The perturbation formulae yield
accurate two-point Pad\'{e} analytical expressions of the tidal
Love numbers of degree $l=2$ and $l=3$, with the errors within
$2.5 \times 10^{-5}$ per cent and $7.5 \times 10^{-5}$ per cent
respectively over the range of polytropic indices $n\in[0,1]$.
Over the entire physical interval of stability, where $n\in[0,3]$,
the errors are still less than $1$ per cent (see
Table~\ref{table:error}).

On the other hand, the scaled moment of inertia (or its multipolar
counterpart) of polytropic stars can  also be found directly from
integrating the density profile resulting from the application of
the scaled delta expansion method.
We see that the I-Love  relation vary less than $0.4$ per cent for
$n \in [0,1]$. However, the I-Love relation and likewise the
multipolar moment-Love relations demonstrate more obvious
dependence on $n$ for $n >1$ (see Table~\ref{table:error}).
Therefore, these relations can be considered as universal only for
sufficiently stiff stars. Thus, our analytical form of the I-Love
relation (or the multipolar moment-Love relations) further supports
the proposal reported in \citet{Sham_2015,ILoveQ_3,ILoveQ_4}.

In addition to the analytical formulae of the I-Love relation and
the multipolar moment-Love relation of polytropic stars in the
Newtonian limit,  we further supplement the universality of these
relations by analysing the perturbative response of the tidal Love
numbers to arbitrary density variations.  We find that the
stationarity of the I-Love relation, as well as the multipolar
moment-Love relation, at the incompressible limit is actually
independent of the exact form of the density variation. In fact,
the stationarity of these relations is due to the cancellation of
the respective variations in the mass, the mass moment, and the
tidal Love number.

The structure and the major findings of the paper are summarized
as follows. In Section \ref{sec:ILQ_intro}, we briefly review the
I-Love-Q relations.
In Section \ref{sec:Polytropes_intro}, we introduce  the
polytropic EOS and  the Lane-Emden equation (LEE) governing the
hydrostatic equilibrium of polytropic stars. In Section
\ref{sec:SDEM}, we outline the scaled delta
expansion method proposed by
\citet{Kenny_SDEM} to perturbatively solve the Lane-Emden equation from the
incompressible limit $n=0$ and the case $n=1$. Such results will
be applied to the perturbation calculations of the tidal Love
numbers and the moments of polytropic stars. In Section
\ref{sec:tidal_Love_numbers}, we review the formulation of tidal
Love numbers in Newtonian gravity, and analyse the effects of a
jump-discontinuity in the density profile on the gravitational
potential. Through balancing the singularities, we define a
modified potential whose spatial derivative remains continuous in
the presence of density discontinuities. In Section
\ref{sec:general_perturbation}, we derive a recursive formula for
the perturbative corrections to the tidal Love numbers from an
arbitrary solvable system.  In particular, we express the
perturbative corrections from the incompressible limit in terms of
the moments and the overlap integrals of the density perturbations
up to the third order. In Section
\ref{sec:polytrope_perturbation}, we apply the results obtained in
the preceding section to the polytropic EOS.  The perturbative
corrections to the tidal Love numbers of degree $l=2$ and $l=3$
are derived, up to the third order about the incompressible limit
$n=0$, and up to the first order about the case $n=1$. Accurate
two-point Pad\'{e} approximants for the tidal Love numbers are
formed from these perturbative results.
In Section \ref{sec:ILoverelations_general},
we apply the perturbation results developed about the
incompressible limit in Section \ref{sec:general_perturbation} to
examine the validity of
the I-Love relation and the multipolar moment-Love relation. 
We show that these relations are stationary with respect to any
density variations away from the incompressible limit. In Section
\ref{sec:ILoverelations_polytropes}, we study the I-Love relation
and the multipolar moment-Love relation of polytropic stars in
greater detail. 
We find that our perturbative results can nicely capture the
behaviour of these relations,
demonstrating the robustness of the perturbative approach developed in the present paper. In
Section \ref{sec:discussion}, we present the conclusion of our
paper and further discuss the physical origin of the universality
of these moment-Love relations.
For reference, we list the scaled delta
expansion method results to the Lane-Emden equation in Appendix
\ref{sec:SDEM_formulae_Theta_only}, 
and argue in Appendix \ref{sec:n5} that the tidal Love numbers
 behaves as $(n-5)^2$ in the limit $n \rightarrow
5$.
In Appendix \ref{sec:formulae_for_lazy_physicist}, accurate approximants
of the radius, tidal Love numbers (deformabilities), mass and moment of
inertia are listed in a self-contained manner.

\section{I-Love-Q relations} \label{sec:ILQ_intro}
The I-Love-Q universal relations first discovered by
\citet{YagiYunes2013, YagiYunes2013Science} connect three
dimensionless (in geometric units where $G=c=1$) physical
quantities of compact stars, namely, the scaled moment of inertia
$\bar{I} \equiv I /M^3$, the dimensionless tidal deformability
$\bar{\lambda} \equiv (2k_2/3) (R/M)^{5}$ and the scaled
spin-induced quadrupole moment $\bar{Q} \equiv M q_{\text{s}}/I^2 \Omega^2$,
where $M, R, I, k_2, q_{\text{s}}$, and $\Omega$ are the mass, radius, moment
of inertia, (quadrupole) tidal Love number, spin-induced
quadrupole moment and the spin (angular) frequency of a star.
Standard procedures evaluating these quantities in the Newtonian
and Einstein gravity have been widely discussed in the literature
\citep[see,
e.g.,][]{tassoulrotationingstars,Hartle_1967,Hartle_1968,Damour_Nagar_2009,Lattimer_2010,YagiYunes2013}.

 The I-Love-Q
relations are universal, i.e., EOS-insensitive, for compact stars.
As shown in Fig. 1 of \citet{YagiYunes2013}, where ${\bar I}$ and
${\bar Q}$ are plotted against ${\bar \lambda}$, the graphs for
compact stars constructed with different EOSs models are close to
each other and the difference between them is usually less than
one per cent level. However, it is worth noting that the EOSs
considered in Fig. 1 of \citet{YagiYunes2013} are either realistic
EOSs of nuclear matter or polytropic EOSs with polytropic indices
less than unity. These popular EOSs for compact stars are all
stiff in the sense that their  adiabatic indices (see Section
\ref{sec:Polytropes_intro} for the definition of adiabatic index)
are larger than 2 (at least in the high density region). To
further investigate the dependence of the I-Love-Q relations on
the stiffness of stellar matter, \citet{Sham_2015} showed in
Fig.~3 of their paper the fractional differences
$\Delta {\bar I} \equiv ( {\bar I} - {\bar I}_{\rm incom} ) /{\bar
I}_{\rm incom}  $
and $\Delta {\bar Q} \equiv ( {\bar Q} - {\bar Q}_{\rm incom}
{)/\bar Q}_{\rm incom}  $, where   ${\bar I}_{\rm incom}$ and
${\bar Q}_{\rm incom}$ are the scaled moments of inertia and the
scaled spin-induced quadrupole moment of incompressible stars,
respectively, for polytropic EOSs with adiabatic indices ranging
from 1.6 to 2.5. They found that both $\Delta {\bar I}$ and
$\Delta {\bar Q}$ increase if the adiabatic index (the
compactness) of a star decreases (increases). As a result, the
I-Love-Q universal relations hold nicely for stiff stars in
Newtonian gravity.

Given the discovery reported in \citet{Sham_2015}, we aim to study
in the present paper the I-Love-Q universal relations for
polytropic stars, whose stiffness can be adjusted continuously, in
Newtonian gravity. Analytic perturbative methods will be developed
to derive accurate approximate formulae, which can work for stars
with a wide range of stiffness, for the I-Love-Q relations. In
particular, we note that $\bar{Q} = \bar{\lambda}/ \bar{I}^2$ in
Newtonian gravity \citep[see,
e.g.,][]{tassoulrotationingstars,YagiYunes2013}. Within the scope
of the present paper, analytical formulae of $k_2$ and $I$ will
suffice to examine the physical nature of the I-Love-Q relations.
Therefore, we shall concentrate on the I-Love relation in the
majority of our paper and show in Section~\ref{sec:discussion} how
the scaled spin-induced quadrupole moment also acquires similar
universal behaviour.
\section{Polytropic Stars} \label{sec:Polytropes_intro}
The EOS of polytropic stars is given by:
\begin{equation}
P(r) = K \rho^{1 + 1/n},
\end{equation}
where $K>0$ and the polytropic index $n \geq 0$ are given
parameters. The stiffness of a polytropic star is usually
characterised by the adiabatic index $\Gamma \equiv
(\rho/P)(dP/d\rho) =1 + 1/n$. For example, polytropic stars with $
\Gamma = \infty$  (i.e., $n=0$) are indeed incompressible.

For a polytropic star with central density $\rho_{c} \equiv
\rho(r=0)$, a length scale $a$ is conventionally introduced:
\begin{equation}
\label{eqn:lengthscale} a = \sqrt{\frac{K (n+1)}{4 \pi G}}
\rho_{c}^{(1-n)/(2n)},
\end{equation}
to define a dimensionless radius $x = r/a$ and the polytrope
function $\theta(x)$:
\begin{equation}
[\theta(x)]^{n} \equiv \frac{\rho(r)}{\rho_{c}} \label{eqn:density_definition}  .
\end{equation}
Under Netwonian gravity and hydrostatic equilibrium condition, the
polytrope function $\theta(x)$ satisfies the Lane-Emden equation (LEE): \citep[see,
e.g.,][]{chandrasekhar1958introduction,Cox,binney}:
\begin{equation}
\label{eqn:polytrope_LEE} \frac{1}{x^2}  \frac{d}{dx} [x^2
\frac{d\theta(x)}{dx}] + [\theta(x)]^{n} = 0,
\end{equation}
as well as the
conventional initial conditions $\theta(0) = 1$ and $\theta'(0) =
0$. Physically speaking, $[\theta(x)]^{n}$ is a measure of the
density distribution. Besides, it is readily shown that
$\theta(x)$ is, up to an additive constant, proportional to the
gravitational potential.

To our knowledge, the
LEE admits exact analytical solution only at $n=0$, $n=1$ and
$n=5$ under the conventional initial conditions mentioned above.
  These closed form solutions could be verified
by direct substitution into the LEE \citep[see][equations (3) -
(5)]{Seidov_LEE}, which are given explicitly as follows:
\begin{align}
\label{eqn:n0thetaexact} & n=0,\quad \theta(x) = 1- \frac{1}{6} x^{2}, \quad \xi = \sqrt{6}, \\
\label{eqn:n1thetaexact} & n=1, \quad \theta(x) = \frac{\sin(x)}{x}, \quad \xi = \pi, \\
\label{eqn:n5thetaexact} & n=5, \quad \theta(x) = \frac{1}{\sqrt{1 + x^{2}/3 }}
\quad \xi = \infty,
\end{align}
where $\xi$ denotes the first zero of the solution $\theta(x)$,
and $\xi = \infty$ means that the solution has no root on the
positive real line. These solutions could be used as the starting
point of perturbation analysis, and  are good checks of numerical
calculations as well.

The total mass $M$ of a polytropic star is given by:
\begin{equation}
M = \rho_{c}^{(3-n)/(2n)} \bigg[ \frac{K(n+1)}{4 \pi G}
\bigg]^{3/2}
   \int_{0}^{\xi} \theta(x)^n 4 \pi x^2
dx.
\end{equation}
Considering the change of the total mass with respect to the
variation of the central density,  we get:
\begin{equation}
\frac{\rho_{c}}{M} \frac{d M}{d \rho_{c}} = \frac{3-n}{2n},
\end{equation}
which is positive when $n<3$ and negative when $n>3$.  It implies
that, polytropes are unstable against radial oscillations when
$n>3$ \citep[see, e.g.,][for the criterion on the stability of
stars]{BWN}. In the present paper we are mainly interested in
stable polytropic stars with $0 \le n \le 3$.

For
polytropic indices other than $n=0$, $n=1$ and $n=5$, there is no
known closed form solution to the LEE. Instead, different
analytical approximate methods to solve the LEE, including
 power series expansion methods \citep[see,
e.g.,][]{Seidov_Series,Hunter,Iacono}, resummation of power series
solutions \citep[see, e.g.,][]{Iacono,Pascual_Pade} and the delta
expansion method (DEM) \citep[see,
e.g.,][]{Bender_delta_expansion,Seidov_LEE,Kenny_SDEM} have been
proposed. In order to evaluate the tidal Love number as well as
the moment of inertia (or other mass moments) of polytropic stars,
we adopt the scaled delta expansion method (SDEM)
\citep{Kenny_SDEM}, which is a variant of the DEM originally
proposed by \citet{Bender_delta_expansion} to solve a wide range
of non-linear problems through expanding the non-linear term in a
series of its power, to solve the LEE and hence the density
profile of polytropic stars. We shall develop proper perturbation
schemes  to determine the tidal Love number and the moment of
inertia (or other mass moments) from the density profile obtained
from the SDEM  in the later part of this paper.

\section{Scaled delta expansion method} \label{sec:SDEM}
Recently, \citet{Kenny_SDEM} have proposed the SDEM based on the
DEM \citep{Bender_delta_expansion,Seidov_LEE} to solve the LEE
perturbatively around $n=0$ and $n=1$, where the LEE admits closed
form solutions in terms of elementary functions. In this section
we briefly review the main idea and the results of the SDEM to
make our paper self-contained.

The spirit of the SDEM is to scale the dimensionless radial
distance $x$ in the LEE \eqref{eqn:polytrope_LEE} in a way that
polytropes of different polytropic indices $n$ share a common
scaled radius at $z=\pi$ through the introduction of a scaling
factor $S(n)$ as follows:
\begin{equation}
\label{eqn:polytrope_def_scale} x = S(n)^{(n-1)/2} z.
\end{equation}
As a result, the scaled polytrope function $\Theta(z) =
\theta(S^{(1-n)/2} z)$ attains its first zero at $z=\pi$ in the
new length scale. Thus, we seek for a solution to the following
scaled LEE (SLEE):
\begin{equation}
\label{eqn:polytrope_SLEE} \frac{1}{z^2}  \frac{d}{dz} [z^2
\frac{d S(n) \Theta(z)}{dz}] + \left[S(n) \Theta(z) \right]^{n} =
0,
\end{equation}
under the three boundary conditions (1) $\Theta(z=0)=1$, (2)
$\Theta'(z=0)=0$, and (3) $\Theta(z=\pi)=0$. The SLEE
\eqref{eqn:polytrope_SLEE} is nicer than the original LEE
\eqref{eqn:polytrope_LEE} because $\Theta(z)>0$ for $0 \le z <
\pi$. As a result, the nonlinear term $\Theta(z)^{n}$ can then be
expanded as a power series in $n$, which  is free of singularities
in the whole physical range $0 \le z < \pi$ and in turn leads to
better accuracy of the ensuing perturbative results.

In general, the SLEE \eqref{eqn:polytrope_SLEE} cannot
simultaneously satisfy the three boundary conditions mentioned
above. For each polytropic index $n$, the additional constraint
allows us to determine the scale factor $S(n)$.  In other words,
the LEE is recast into an eigenvalue problem, with the scale
factor $S$ being the eigenvalue and the scaled polytrope function
$\Theta(z)$ being the eigenfunction.

As the SLEE admits closed form solution of finite radius at $n=0$
and $n=1$, the scale factor $S$ and the polytrope function
$\Theta(z)$ are solved at these points by expanding them into
perturbation series in $n$ and $n-1$, respectively, as follows:
\begin{align}
\nonumber \Theta(z) = & \Theta_{p}^{(0)}(z) + (n-p) \Theta_{p}^{(1)}(z) + (n-p)^2 \Theta_{p}^{(2)}(z) \\
& + (n-p)^3 \Theta_{p}^{(3)}(z) + O[(n-p)^4],
\end{align}
\begin{align}
\nonumber S(n) = & S_{p}^{(0)} + (n-p) S_{p}^{(1)} + (n-p)^2 S_{p}^{(2)} \\
\label{eqn:Sn_perturbative_expansion} & + (n-p)^3 S_{p}^{(3)} + O[(n-p)^4],
\end{align}
where $p=0$ or $1$ denotes the perturbation centre, $n-p$ is
considered as the perturbation parameter, and $\Theta_{p}^{(j)}$
and $S_{p}^{(j)}$ respectively denote the $j$th order perturbation
correction of the scaled polytrope function $\Theta(z)$ and scale
factor $S$ about the perturbation centre $p$.  The perturbative
corrections to the scale polytrope function $\Theta(z)$ have been
solved up to the second order at $n=0$, and the first order at
$n=1$ \citep{Kenny_SDEM}, and are summarized in Appendix
\ref{sec:SDEM_formulae_Theta_only}.

The SDEM approach is particularly useful in our present context.
Firstly, as will be shown in Section~\ref{sec:tidal_Love_numbers},
the tidal Love numbers  are independent of the length scale. In
the SDEM approach, finite polytropes of different physical radii
are scaled to the same radius at $\pi$ in the scaled length
variable $z=\pi$ via the scale factor $S(n)$. In the determination
of the tidal Love number, the length scale factor $S(n)$
associated with the $n$-dependence of the physical radius
completely disappears, so there is no need to look for the
solution to $S(n)$ in equation
(\ref{eqn:Sn_perturbative_expansion}).  Only the density scaled to
a common radius at $z=\pi$ has to be considered. Secondly, the
SDEM approach determines the corrections of the scaled polytrope
function perturbatively from the exact solutions of the SLEE
available  at $n=0$ (i.e., the incompressible limit) and $n=1$.
These perturbative corrections to the scaled polytrope function
straightforwardly determine  the density perturbation
$\rho_{1}(r)$ (see equation \eqref{density}) and consequently form
the input to the perturbative analyses developed in the following
sections.

\section{Tidal Love number} \label{sec:tidal_Love_numbers}
Traditionally, multipolar tidal Love numbers can be obtained by
solving the Poisson's equation that governs the spatial variation of
the gravitation potential inside a star. In general, whenever the
density of a star encounters a jump discontinuity (e.g., the
density discontinuity at the surface of a QS), there is an
associated jump discontinuity in the derivative of the gravitation
potential, which could render perturbative expansion inapplicable.
To remedy this problem, in the subsequent discussion a modified
potential, which is continuous everywhere inside the star,
is introduced to replace the traditional gravitation potential. We
show that the multipolar tidal Love numbers can be found by
solving the governing equation of such a modified potential.

\subsection{Definition}
Consider a spherically symmetric star of radius $R$, density
profile $\rho(r)$ and pressure profile $P(r)$ under the influence
of an axially symmetric  external multipole tidal field $H_{\rm
ext} \equiv A_l r^l P_l (\cos \theta)$, where $A_l$ is a constant,
$P_l (\cos \theta)$ is the standard Legendre polynomial of order
$l$, and $l=2,3,4,\ldots$. Inside the star, the Eulerian change of
the Newtonian gravitational potential $H(r)$  satisfies the
standard Poisson's equation \citep[see, e.g.,][and references
therein]{Brooker_1955,Beuthe_2015,Lattimer_2010,
Damour_Nagar_2009}:
\begin{equation}
\label{eqn:H_definition}
H''(r) + \frac{2}{r} H'(r) - \bigg[ \frac{l(l+1)}{r^2} - 4 \pi G \rho(r) \frac{d \rho}{dP} \bigg] H(r) = 0,
\end{equation}
where $G$ is the constant of universal gravitation and the term
$\rho (d \rho/dP) H $ is the Eulerian change in the mass density
in response to the tidal field.  On the other hand, outside the
star $H(r)= H_{\rm ext} + q_l r^{-(l+1)} P_l (\cos \theta)$, and
the multipole moment $q_l$ of the star induced by the tidal field
is defined accordingly. Equation \eqref{eqn:H_definition} is
solved under the condition that $H(r)$ is continuous at $r=R$.


The tidal Love number $k_{l}$ of degree $l$ ($l=2,3,4,...$), also
known as the constants of apsidal motion, is given by half of the
limiting ratio of the  gravitational potential due to the so
induced multipole moment $q_l$ to the applied gravitational
potential evaluated at surface of the star \citep[see,
e.g.,][]{Brooker_1955,Lattimer_2010, Damour_Nagar_2009,
Beuthe_2015}, i.e.,
\begin{equation}\label{eqn:kldef}
k_{l} \equiv \frac{q_l}{2 A_l R^{2l+1}}  .
\end{equation}
 In the I-Love-Q relations, the quadrupolar  tidal Love number
$k_{2}$  is considered specifically
\citep{YagiYunes2013,YagiYunes2013Science}.

It follows directly from \eqref{eqn:H_definition} that $k_{l}$ can
be found from the ratio $H'(R^{+})/H(R)$ as follows:
\begin{equation}
\label{eqn:kl_definition} k_{l} = \frac{l - [R H'(R^{+})/H(R)]}{2
\{ l+1+ [R H'(R^{+})/H(R)] \}},
\end{equation}
where, in general, $r^{\pm} \equiv r \pm \delta$ in the limit of
$\delta \rightarrow 0^+$. It is straightforward to show  from
equation \eqref{eqn:H_definition} that as the result of a density
jump of $\Delta \rho \equiv \rho(r_*^+)-\rho(r_*^-)$ across $r =
r_{*}$, $H'(r)$ correspondingly develops a discontinuity of $4 \pi
r_{*}^2 H(r_{*}) \Delta \rho / m(r_{*})$ \citep[see Section III
of][for a discussion on this issue]{Lattimer_2010}, where $m(r)$
is the mass enclosed within the radius $r$:
\begin{equation}
m(r) = \int_{0}^{r} \rho(r) 4 \pi r^2 dr.
\end{equation}
Therefore, for stars with non-vanishing surface mass density
$\rho(R^{-})$, e.g., incompressible stars and QSs, equation
\eqref{eqn:kl_definition} has to be modified as: 
\begin{equation}
\label{eqn:kl_interior} k_{l} = \frac{l - [R H'(R^{-})/H(R) - 4
\pi R^3 \rho(R^{-}) / m(R)]}{2 \{ l+1+ (R H'(R^{-})/H(R) - 4 \pi
R^3 \rho(R^{-}) / m(R)) \}},
\end{equation}
if the ratio $H'(R^{-})/H(R)$, instead of $H'(R^{+})/H(R)$, is
used to evaluate $k_{l}$
 \citep[see, e.g.,][]{Lattimer_2010}.

 On the other hand,
as the tidal Love numbers $k_{l}$ depends only on the ratio $ R
H'(R) / H(R)$, it is customary to  recast equation
(\ref{eqn:H_definition}) into a first-order nonlinear Ricatti
differential equation \citep[see, e.g.,][]{Lattimer_2010}:
\begin{equation}
\label{eqn:conventional_Y} Y'(r) + \frac{1}{r} Y(r) + \frac{1}{r}
Y(r)^2 = \frac{l(l+1)}{r} + \frac{4 \pi r^3}{m(r)} \frac{d \rho}{d
r},
\end{equation}
where
\begin{equation} \label{LDP}
 Y(r) \equiv \frac{d \ln H}{d \ln r} =r \frac{H'(r)}{H(r)}
\end{equation}
 is the logarithmic derivative of the potential $H(r)$  (LDP)
satisfying the initial condition $Y(0) = l$.

A couple of remarks are as follows. Firstly, equation
(\ref{eqn:conventional_Y}) is invariant under length (or mass)
scale transformation.
Hence, tidal Love numbers are independent of the size and the
density scale of the stellar object, i.e., two stars of density
profiles $\rho(r)$ and $\alpha \rho(\beta r)$ have the same tidal
Love numbers, for arbitrary $\alpha >0$ and $\beta >0$. As a
result, there is no need for us to explicitly determine the
scaling factor $S(n)$ introduced in the SDEM because $k_l$ is
independent of $S(n)$.

 Secondly, it is easy to see that both equations \eqref{eqn:H_definition} and
 (\ref{eqn:conventional_Y}) could contain divergent terms in some situations.
 For example,
for polytropes of polytropic index $n \ne 0 $, it is well known
that near the stellar surface, the pressure $P(r)$, the density
$\rho(r)$ and $d \rho(r)/dr $  behave as $(R - r)^{n+1}$, $(R -
r)^{n}$ and $(R - r)^{n-1}$, respectively. As mentioned
previously, in equation \eqref{eqn:H_definition}, the induced
change in the mass density is proportional to $\rho (d \rho/ d p)
\propto (R - r)^{n-1}$. Similarly, on the RHS of equation
(\ref{eqn:conventional_Y}), there is a term proportional to $d
\rho(r)/dr \propto (R - r)^{n-1}$. These terms are obviously
divergent as $r$ approaches $R$ for polytropes with $n \in (0,1)$.
On the other hand, they are also problematic for stars with
density discontinuities or rapidly varying density profile.
In fact,  $d \rho(r)/dr$ is actually proportional to a
delta-function at a density discontinuity. These diverging terms
in equations \eqref{eqn:H_definition} and
(\ref{eqn:conventional_Y}) are likely to hamper both analytical
perturbative as well as numerical calculations \citep[see the
relevant discussion in][]{Lattimer_2010}. In order to get rid of
these troublesome terms, we propose in the following a modified
potential whose governing equation and the associated Ricatti
equation are free of divergence.

\subsection{Modified potential}
To remedy the problem of divergence in equations
(\ref{eqn:H_definition}) and (\ref{eqn:conventional_Y}) as
mentioned above, we consider the modified potential $h(r)$ defined
by:
\begin{equation}
h(r) = H(r)/m(r),
\end{equation}
and obtain the differential equation governing $h(r)$:
\begin{equation}
\label{eqn:h} h''(r) + \left[\frac{2}{r} + \frac{8 \pi r^2
\rho(r)}{m(r)} \right] h'(r) + \bigg[\frac{16 \pi r \rho(r)}{m(r)}
-\frac{l(l+1)}{r^2} \bigg] h(r) = 0.
\end{equation}
The proper boundary condition for this modified potential is $h(r)
\sim r^{l-3}$ as $r \rightarrow 0$. It is obvious that this new
field equation is
divergence-free (except at the origin). 
As a result, while $H'(r)$ is discontinuous across density
discontinuities where $d \rho /d P$ is infinite, $h'(r)$ is
continuous across such points.

Moreover, it can be shown that the logarithmic derivative of the
modified potential (LDMP), $y(r) \equiv d \ln h /d \ln r =  r
h'(r)/h(r)$, is  related to the LDP, $Y(r)$, through the following
equation:
\begin{equation}
\label{eqn:y_def} y(r) = Y(r) - \frac{4 \pi r^3 \rho(r)}{m(r)},
\end{equation}
and the governing equation of $y(r)$ can be obtained from equation
(\ref{eqn:h}):
\begin{equation}
\label{eqn:y} y'(r) + \bigg[\frac{1}{r} + \frac{8 \pi r^2
\rho(r)}{m(r)} \bigg] y(r)  + \frac{1}{r} {y(r)}^2 = \frac{l(l+1)}{r} - \frac{16 \pi r^2
\rho(r)}{m(r)},
\end{equation}
which is to be solved under the initial condition $y(0) = l - 3$.
It follows directly from equations \eqref{eqn:kl_interior} and
\eqref{eqn:y_def} that the tidal Love numbers, $k_{l}$, depends
only on the LDMP, $y(r)$:
\begin{equation}
\label{eqn:kl_yl} k_{l} = \frac{l - y(R)}{2 [l+1+ y(R)] },
\end{equation}
As the derivative of the density profile no longer appears
explicitly in the governing equation of $y(r)$, $y(r)$ is amenable
to perturbative calculations even for cases with $n \in [0,1)$. We
shall see in the subsequent discussion equation \eqref{eqn:y}
indeed opens up the possibility of perturbative analysis from the
incompressible limit (i.e., $n=0$). Otherwise, for a slight
perturbation of the density profile from the incompressible limit,
the derivative of the change in the density profile appearing in
\eqref{eqn:conventional_Y}, in general, dominates the derivative
of the original constant density profile, and precludes the
perturbative analysis of $Y(r)$.

The delicate balance between the discontinuities in the two terms
$Y(r)$ and $4 \pi r^3 \rho(r)/m(r)$ appearing in the definition of
$y(r)$ echoes the relationship between the electric field and
electric displacement field in the theory of electrostatics in
dielectrc media.  While the electric field is discontinuous across
the surface of a polarized material, the electric displacement
takes into account of the contributions due to polarization in a
way that the electric displacement is continuous \citep[see,
e.g.,][]{JacksonEM}.

It is interesting to note that the Ricatti equation \eqref{eqn:y}
originally derived from the consideration of the Eulerian change
in the gravitational  potential $H$ is analogous to the
Clairaut-Radau equation in the traditional theory of tides
\citep[see, e.g.,][]{Brooker_1955,tassoulrotationingstars}, which
was obtained by monitoring the deformation of the equivpotential
surfaces inside a star upon the influence of a tidal field. In
fact, the solutions of these two equations are related through the
linear transformation $y(r) \rightarrow y(r) - l + 3$.


\subsection{Exactly solvable models}
For the case of polytropic stars, equation \eqref{eqn:y} can be
exactly solved for two different cases, namely, $n=0$ and $n=1$.
For the case $n=0$ (i.e., incompressible stars),  the governing
equation  (\ref{eqn:y}) reduces to, for $r < R$:
\begin{equation}
y'(r) + \frac{7}{r} y(r) + \frac{1}{r} y(r)^2 = \frac{l^2 + l -12}{r},
\end{equation}
which is solved under the initial condition $y(0) = l-3$. By
inspection, one can see that, for $r < R$:
\begin{equation}
y(r) = l-3.
\end{equation}
As $y$ is continuous at the stellar surface, 
the tidal Love number $k_{l}$ of an incompressible star can be
obtained from equation (\ref{eqn:kl_yl}):
\begin{equation}
k_{l} = \frac{3}{4(l-1)}.
\end{equation}

For the case with $n=1$, for $r<R$, the tidal field equation
$H(r)$ is the standard spherical Bessel equation:
\begin{equation}
H''(r) + \frac{2}{r} H'(r) - \bigg[ \frac{l(l+1)}{r^2} - \frac{1}{a^2} \bigg]H(r) =0.
\end{equation}
The solution that is non-singular at the origin is the standard
spherical Bessel function,
\begin{equation}
H(r) \propto j_{l}(r/a),
\end{equation}
  for $r \le R$. Besides, we also have:
\begin{equation}
\label{eqn:n1rho0} \rho(r) = \rho_{c} \frac{\sin(r/a)}{r/a}.
\end{equation}
Therefore, we can determine the $y_{l}(r)$ by equation 
(\ref{eqn:y_def}):
\begin{equation}
\label{eqn:n1y0}
y(r) = l - \frac{r j_{l+1}(r/a)}{a j_l(r/a)} + \frac{r^2 \sin(r/a)}{a r \cos(r/a)- a^2 \sin(r/a)}.
\end{equation}
Hence, evaluating at the stellar surface at $R = a \pi$, by equation (\ref{eqn:kl_yl}), we have:
\begin{equation}
k_{l} = \frac{\pi}{2} \frac{j_{l+1}(\pi)}{(2l+1)j_{l}(\pi) - \pi j_{l+1}(\pi)}.
\end{equation}

Although the LEE also admits exact solution at $n=5$, neither
$H(r)$ nor $h(r)$ can be found analytically for the case $n=5$, at
least  to our knowledge. However, it can be shown that $k_l$ is
proportional to $(n-5)^2$ for $n$ close to 5. Hence, $k_l$
vanishes identically for polytropes with $n=5$ as a consequence of
their infinite spatial extent. These results are useful in
improving the accuracy of the perturbative formula (to be
developed in the subsequent discussion) for $k_l$. In order not to
distract the readers' attention from the main theme of the present
paper, instead of working out a detailed proof for such asymptotic
behaviour here, we move onward to develop a perturbative theory
for the tidal Love number in the following section. However, we
will briefly outline the proof in Appendix~\ref{sec:n5}.

\section{Perturbative expansion of Tidal Love Numbers} \label{sec:general_perturbation}
The removal of the explicit dependence on the derivative of
density profile in the governing equation (\ref{eqn:y}) allows
perturbative analysis of the tidal Love numbers.  We first derive
a recursion formula for the perturbative corrections of the LDMP,
$y(r)$, due to variation of the density profile from an arbitrary
exactly solvable state.  Then, we consider an  incompressible star
as the unperturbed system and find that the perturbative
corrections to the tidal Love numbers are expressible  in terms of
the moments of the density perturbations.

\subsection{General cases}
We derive the effects on the modified potential  due to a density
perturbation $\epsilon \rho_{1}(r)$, where $\epsilon$ is a
 parameter measuring the strength of perturbation, on an exactly
solvable star with a given
 density distribution $\rho_{0}(r)$ and correspondingly a known solution $y_{0}(r)$ to \eqref{eqn:y}.
As a result, the density profile $\rho(r)$ and the 
LDMP, 
$y(r)$, can be
expressed respectively as:
\begin{align}
& \rho(r) = \rho_{0}(r) + \epsilon \rho_{1}(r),  \label{density} \\
& y(r) = y_{0}(r) + \epsilon y_{1}(r) + \epsilon^2 y_{2}(r)
+\ldots,
\end{align}
where $\epsilon^j y_{j}(r)$ is the $j$th order perturbative
correction to $y(r)$ due to the density perturbation $\epsilon
\rho_{1}(r)$. For notational convenience, we write:
\begin{equation}
m(r) = m_{0}(r) + \epsilon m_{1}(r) = \int_{0}^{r} \rho_{0}(t) 4 \pi t^2 dt + \epsilon \int_{0}^{r} \rho_{1}(t) 4 \pi t^2 dt.
\end{equation}

The zeroth order and the higher order equations for $j\geq1$ read:
\begin{equation}
\frac{d y_{0}}{d r} + \frac{1}{r} y_{0} + \frac{1}{r} y_{0}{}^2 +
\frac{8 \pi r^2 \rho_{0}}{m_{0}}  y_{0} = \frac{l(l+1)}{r} -
\frac{16 \pi r^2 \rho_{0}}{m_{0}},
\end{equation}
\begin{align}
\nonumber \frac{d y_{j}}{d r} + \bigg( \frac{2 y_{0} + 1}{r} + \frac{8 \pi r^2 \rho_{0}}{m_{0}} \bigg) y_{j}
= &  8 \pi r^2\bigg( \frac{ \rho_{1}}{m_{1}} - \frac{ \rho_{0}}{m_{0}}  \bigg) (y_{0} + 2)  \\
& + g_{j},
\end{align}
where $g_{1}(r) = 0$, and for $j \geq 2$:
\begin{equation}
g_{j}(r) = \sum_{k=1}^{j-1} \bigg[ 8 \pi r^2\bigg( \frac{
\rho_{1}}{m_{1}} - \frac{ \rho_{0}}{m_{0}} \bigg)
\big(-\frac{m_{1}}{m_{0}}\big)^{k} - \frac{y_{k}}{r}
\bigg]y_{j-k},
\end{equation}
which are to be solved under the initial conditions $y_{j}(0) = (l
- 3) \delta_{0j}$. The zeroth order equation, by assumption, is
exactly solvable.  The higher order equations are first order
linear differential equation, so that the $j$th order solution,
$y_{j}$, is expressible in terms of the lower order solutions,
$\{y_{0}, y_{1}, \ldots, y_{j-1}\}$, by quadrature:
\begin{align}
\nonumber y_{j}(r) = & \frac{1}{m_{0}(r){}^2} \exp \bigg( - \int_{0}^{r} \frac{2 y_{0}(t) + 1}{t} dt \bigg) \\
\nonumber & \times \int_{0}^{r} m_{0}(s){}^{2} \exp \bigg( \int_{0}^{s} \frac{2 y_{0}(t) + 1}{t} dt \bigg) \\
\label{eqn:yj_solution} & \times \bigg\{ 8 \pi s^2 \bigg( \frac{\rho_{1}(s)}{m_{1}(s)} -
\frac{\rho_{0}(s)}{m_{0}(s)}  \bigg) [y_{0}(s) + 2] +
g_{j}(s) \bigg\} ds.
\end{align}

\subsection{Incompressible stars} \label{sec:incompressible_perturbation_general}
As mentioned previously, the case of incompressible stars  is one
of the few exactly solvable models whose tidal Love number can be
found analytically. Here we consider an unperturbed star of
constant density $\rho_{0}$. Then, we have, for $r < R$:
\begin{align}
& m_{0}(r) = \frac{4 \pi}{3} \rho_{0} r^3, \\
& y_0(r) = l-3.
\end{align}
In this case, equation (\ref{eqn:yj_solution}) could be
simplified, and we shall see that the perturbative corrections
$y_{j}$ are expressible in terms of the moments of the density
perturbation $\rho_1(r)$. For notational convenience, we denote
the dimensionless $k$th moment $\mu_{k}(r)$ of the dimensionless
density perturbation by:
\begin{equation}
\mu_{k}(r) = \frac{1}{r^{k+3}} \int_{0}^{r}  \frac{\rho_{1}(t)}{\rho_{0}} t^{k+2} dt.
\end{equation}

Solving equation (\ref{eqn:yj_solution}) recursively for
$y_{0}^{(0)}(r)$, $y_{0}^{(1)}(r)$, $y_{0}^{(2)}(r)$, and
$y_{0}^{(3)}(r)$, we obtain:
\begin{align}
\nonumber y(r) = & l-3 + \epsilon \bigg[9 \mu_{0} - 3(2l+1)\mu_{2l-2} \bigg] \\
\nonumber & + \epsilon^2 \bigg\{-27 \mu_{0}{}^2 + 9(2l+1) (\mu_{2l-2})^2 + 18(l-1) \mathcal{C}[\mu_{0}]\bigg\} \\
\nonumber & +  \epsilon^3 \bigg\{ 81 \mu_{0}{}^3 - 27(2l+1)(\mu_{2l-2})^3 - 108(l-1) \mu_{2l-2} \mathcal{C}[\mu_{0}] \\
\nonumber &  + 27(-2l+5) \mathcal{C}[(\mu_{0})^2]- 162 \mathcal{C}[\mu_{0} \mu_{2l-2}] \\
\label{eqn:y_solution_inc} &  + 27(2l +
1)\mathcal{C}[(\mu_{2l-2})^2]\bigg\} + O[\epsilon^4],
\end{align}
where $\mathcal{C}$ denotes the dimensionless overlap integral with $\rho_{1}(t) t^{2l}$:
\begin{equation}
\label{eqn:C_definition} \mathcal{C}[f](r) = \frac{1}{r^{2l+1}}
\int_{0}^{r} f(t) \frac{\rho_{1}(t)}{\rho_{0}} t^{2l} dt,
\end{equation}
and the $r$-dependence of $\mu_{0}$, $\mu_{2l-2}$, $
\mathcal{C}[(\mu_{0})^2]$, $\mathcal{C}[\mu_{0} \mu_{2l-2}]$, and
$\mathcal{C}[(\mu_{2l-2})^2]$ are suppressed. By equation
(\ref{eqn:kl_yl}),  the tidal Love number of degree $l$ is
evaluated perturbatively  from the incompressible limit and given
up to the third-order in $\epsilon$ by:
\begin{align}
\nonumber k_{l} = & \frac{3}{4 (l-1)} + \frac{3 \epsilon (2 l+1)}{8 (l-1)^2} \bigg[ -3\mu_{0} + (2l+1)\mu_{2l-2} \bigg] \\
\nonumber & + \frac{9 \epsilon^2 (2 l+1)}{16 (l-1)^3} \bigg\{ 3 (2l + 1) \big( \mu_{0} -\mu_{2l-2} \big)^2 - 4(l-1)^2  \mathcal{C}[\mu_{0}] \bigg\} \\
\nonumber & + \frac{27 \epsilon^3 (2 l+1)}{32 (l-1)^4} \bigg\{ 3 (2l+1) \big(\mu_{0} -\mu_{2l-2} \big)^2 \\
\nonumber & \times \big[ -(2l+1) \mu_{0} + 3 \mu_{2l-2} \big] + 24 (l-1)^2 \mathcal{C}[\mu_{0}] \big( \mu_{0} - \mu_{2l-2} \big) \\
\nonumber & + 4 (l-1)^2 \big[(2l-5) \mathcal{C}[(\mu_{0})^2] + 6 \mathcal{C}[\mu_{0} \mu_{2l-2}] \\
\label{eqn:kl_solution_inc} &  - (2 l+1) \mathcal{C}[(\mu_{2l-2})^2] \big]   \bigg\}  + O[\epsilon^4],
\end{align}
where all the moments 
and the overlap integrals 
appearing in RHS of equation \eqref{eqn:kl_solution_inc} are
evaluated at the stellar surface $r=R$.

We hint that the perturbation solution offers an analytical
understanding the I-Love relation.  The first order perturbative
response of the tidal Love number is proportional to the zeroth
moment, and the $(2l-2)$th moment of the density perturbation.  In
the special case when $l=2$, the moments correspond to the
variations of the mass and the moment of inertia of the stellar
structure.  We shall see in Section
\ref{sec:ILoverelations_general} that the tidal Love number, the
mass and the moment of inertia are combined in the I-Love
relation, so that the perturbation responses cancel exactly in the
first order.

\section{Tidal Love Numbers of Polytropic stars} \label{sec:polytrope_perturbation}
In this section we shall use the SDEM results of the SLEE
\citep{Kenny_SDEM} and the perturbative expansion of the tidal
Love number derived in the preceding section to find the
perturbative series of the polytropic tidal Love numbers of degree
$l=2$ and $l=3$, up to the third order about the incompressible
limit $n=0$, and up to the first order about the case $n=1$. With
these two series derived respectively at $n=0$ and $n=1$,  we form
a two point Pad\'{e} approximants of $k_{2}$ and $k_{3}$, whose
errors are within $0.39$ per cent and $0.93$ per cent respectively
for stable polytropes for $n\in[0,3]$. Between the two
perturbation centres $n\in[0,1]$, the errors are $2.5 \times
10^{-5} $ per cent and $7.5 \times 10^{-5} $ per cent respectively
(see Table~\ref{table:error}). In the following we provide the
technical details of our calculations.

\subsection{$n \approx 0$ case}
The density of a polytrope is proportional to $\Theta^n$, so the
density perturbation away from the incompressible limit is given
up to the third order in $n$ by \citep{Kenny_SDEM}:
\begin{align}
\nonumber \varepsilon \rho_{1}(z; n=0) = & n \rho_{c} \bigg[ \ln \Theta_{0}^{(0)} + n \left(\frac{1}{2} \ln ^2 \Theta_{0}^{(0)} + \frac{\Theta_{0}^{(1)}}{\Theta_{0}^{(0)}} \right) \\
\nonumber & + n^2 \bigg(\frac{1}{6} \ln ^3 \Theta_{0}^{(0)}  + \frac{\Theta_{0}^{(1)} \ln \Theta_{0}^{(0)}}{\Theta_{0}^{(0)}} \\
& - \frac{\Theta_{0}^{(1)}{}^2}{2
\Theta_{0}^{(0)}{}^2}+\frac{\Theta_{0}^{(2)}}{\Theta_{0}^{(0)}}\bigg)
\bigg] + O[n^4],
\end{align}
where the perturbation corrections $\Theta_{0}^{(j)}$ for $j=0$,
$1$ or $2$ are obtained by \citet{Kenny_SDEM} and listed in
Appendix~\ref{sec:SDEM_formulae_Theta_only} for reference. In this
case we shall let  $n$ be the perturbation parameter $\varepsilon$
from the incompressible limit and use equation
\eqref{eqn:kl_solution_inc} to evaluate
$k_{2}$ and $k_{3}$,
yielding the results:
\begin{align}
\nonumber k_{2} & = \frac{3}{4} -\frac{3}{4} n + \bigg( \frac{766}{75} - \pi ^2 \bigg) n^2 + n^3 \bigg[ - \frac{471797}{1875} + \frac{173}{25}\pi^2 \\
\nonumber & \quad + (-\frac{1087}{75} + \frac{2}{5} \pi^2)\ln2 + \frac{216}{25} \ln^2 2  - \frac{12}{5} \ln^3 2 + 156 \zeta (3) \\
\nonumber & \quad - \frac{135}{8} J(2,2;4) + \frac{405}{4} J(2,4;4) - \frac{675}{8} J(4,4;4) \bigg] + O[n^4]\\
& \label{eqn:polytrope_k2n0} \approx 0.75 - 0.75 n + 0.343728932 n^2 - 0.10794273 n^3 + O[n^4], \\
\nonumber k_{3} & = \frac{3}{8} -\frac{9}{20} n + \bigg( \frac{37531}{4900} - \frac{3 \pi ^2}{4} \bigg)n^2 + n^3 \bigg[ -\frac{1141420459}{5402250} \\
\nonumber & \quad + \frac{11141}{1960} \pi^2 + ( -\frac{289253}{25725}+\frac{2}{7} \pi ^2) \ln 2 + \frac{1608}{245} \ln^2 2 \\
\nonumber & \quad - \frac{12}{7}  \ln^3 2 + 132 \zeta (3) + \frac{189}{32} J(2,2;6) + \frac{567}{16} J(2,6;6) \\
\nonumber & \quad -\frac{1323}{32} J(6,6;6) \bigg] + O[n^4] \\
& \label{eqn:polytrope_k3n0} \approx 0.375 - 0.45 n + 0.257184454
n^2 - 0.103284215 n^3 + O[n^4],
\end{align}
where $\zeta(s)$ is  the Riemann zeta function defined by the
infinite series \citep[see, e.g.,][]{Math_handbook}:
\begin{equation}
\zeta(s) = \sum_{k=1}^{\infty} \frac{1}{k^{s}}
\end{equation}
for $s>1$, and $J(p,q;r)$ is a third-order interaction integral
defined to be:
\begin{align}
\nonumber J(p,q;r) = & \int_{0}^{1} \bigg[ \frac{1}{t^{p+1}} \int_{0}^{t} \ln(1-s^2) s^p ds  \bigg] \bigg[ \frac{1}{t^{q+1}} \int_{0}^{t} \ln(1-s^2) s^q ds  \bigg]  \\
& \times \ln(1-t^2) t^{r} dt.
\end{align}
We note that $J(p,q;r) = J(q,p;r)$, and these integrals are caused by the third order interaction integrals $\mathcal{C}[(\mu_{0})^2]$, $\mathcal{C}[\mu_{0} \mu_{2l-2}]$ and $\mathcal{C}[(\mu_{2l-2})^2]$.

\subsection{$n \approx 1$ case}
Next, we apply the SDEM results developed about $n=1$  to derive
the approximate expressions for $k_{2}$ and $k_{3}$ that are valid
for $n \approx 1$.
Polytropic stars with $n=1$ are  exactly solvable with
$\rho_{0}(r)$ and $y_{0}(r)$  given by equations
(\ref{eqn:n1rho0}) and (\ref{eqn:n1y0}). On the other hand, for
polytropic stars with $n \approx 1$, the density perturbation
is given up to the first-order in $n-1$:
\begin{align}
\varepsilon \rho_{1}(z; n=1) & = (n-1) \rho_{c} \bigg(
\Theta_{1}^{(1)}+ \Theta_{1}^{(0)} \ln \Theta_{1}^{(0)} \bigg) +
O[(n-1)^2],
\end{align}
where the explicit forms of the perturbation corrections
$\Theta_{1}^{(j)}$ for $j=0,1$ can be found in
Appendix~\ref{sec:SDEM_formulae_Theta_only} \citep[see
also][]{Kenny_SDEM}.

Considering $ n-1$ as the perturbative parameter $\varepsilon$ in
equation (\ref{eqn:yj_solution}), and using equations
(\ref{eqn:kl_yl}) and (\ref{eqn:yj_solution}), we obtain:
\begin{align}
\nonumber k_{2} & = \frac{15 - \pi^2}{2 \pi ^2} - \frac{45}{2\pi^4} y_{1}(z=\pi;l=2) (n-1) \\
& \label{eqn:polytrope_k2n1} \approx 0.259908877 - 0.295642768 (n-1) + O[(n-1)^2], \\
\nonumber k_{3} & = \frac{105 - 10 \pi^2}{6 \pi ^2} -\frac{7 \left(15-\pi ^2\right)^2}{18 \pi ^4}y_{1}(z=\pi;l=3) (n-1) \\
& \label{eqn:polytrope_k3n1} \approx 0.106454047 -0.142830875 (n-1) + O[(n-1)^2],
\end{align}
where $y_{1}(z;l)$ for $l=2$ and $l=3$ are defined in equation (\ref{eqn:yj_solution}).

\subsection{Two-point approximations}
We form two-point Pad\'{e} approximants of the tidal Love numbers
$k_{2}$ and $k_{3}$, which are valid for stable polytropic stars
with $n \in [0,3]$. 
Such Pad\'{e} approximants, denoted by $k_{2,\text{tp}}$ and
$k_{3,\text{tp}}$, are in the form of:
\begin{align}
\label{eqn:klPade}k_{l,\text{tp}}(n) = (5-n)^3 \frac{a_{1} + n
a_{2} + n^2 a_{3} + n^3 a_{4}}{1000 + n a_{5} + n^2 a_{6}},
\end{align}
where $l=2$ or $3$, the constants $a_{1}, a_{2}, \ldots, a_{6}$
are determined analytically from the perturbation solutions in
equations (\ref{eqn:polytrope_k2n0}), (\ref{eqn:polytrope_k3n0}),
(\ref{eqn:polytrope_k2n1}) and (\ref{eqn:polytrope_k3n1}) by
applying the two-point Pad\'{e} approximation scheme
\citep{baker1975essentials}, and their numerical values are shown
in Table~\ref{table:1}.
In the above expression, we have also incorporated an empirical
$(5-n)^3$ factor to match the fact that both $k_{2}$ and $k_{3}$
vanish in the limit $n \rightarrow 5$.
Readers may refer to Appendix \ref{sec:n5} for more discussion of
the empirical factor $(5-n)^{3}$.
\begin{table}
\caption{\label{table:1} The numerical values of the constants
$a_{1}, a_{2}, \ldots, a_{6}$ appearing in \eqref{eqn:klPade}.}
\centering
 \begin{tabular}{|l |l | l |}
    \hline
            & $l=2$             & $l=3$ \\
    \hline
    $a_{1}$ & $6$                   & $3$                   \\
    $a_{2}$ & $0.0701474990944$     & $-0.0540432029285$   \\
    $a_{3}$ & $-0.292565896945$     & $-0.248516990271$     \\
    $a_{4}$ & $0.0269208110373$     & $0.0346247187561$      \\
    $a_{5}$ & $411.691249849$       & $581.985599024$       \\
    $a_{6}$ & $17.6102741235$       & $60.5271512307$       \\
    \hline

 \end{tabular}
\end{table}

\subsection{Numerical results}
\begin{figure}
    \includegraphics[width = 0.45 \textwidth]{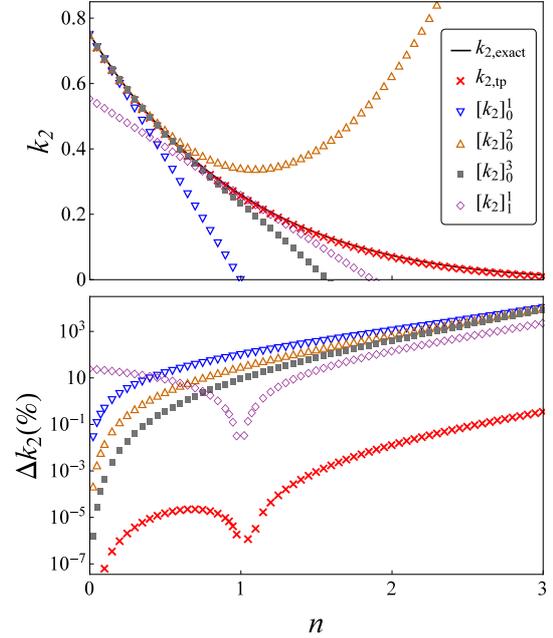}
    \caption{\label{fig:k2_convergence_accuracy}
    In the upper panel, the value of the tidal Love number of degree $l=2$, $k_{2}$, is plotted for
    the numerically exact solution $k_{2,\text{exact}}$ (black line),
    the two-point Pad\'{e} approximant $k_{2,\text{tp}}$ (red cross) in equation (\ref{eqn:klPade}),
    the approximants about $n=0$, $[k_{2}]^{j}_{0}$ in equation (\ref{eqn:polytrope_k2n0})
    for $j=1$ (blue triangle), $j=2$ (orange triangle) and $j=3$ (grey square), and
    the first-order approximation about $n=1$, $[k_{2}]^{1}_{0}$ in equation (\ref{eqn:polytrope_k2n1}) (purple diamond).
    The associated error plot is shown in the lower panel, where
    $\Delta k_{2} = |k_{2} - k_{2,\text{exact}}|/k_{2,\text{exact}} \times 100\%$
    represents the percentage error of the approximant.
    }
\end{figure}

\begin{figure}
    \includegraphics[width = 0.45 \textwidth]{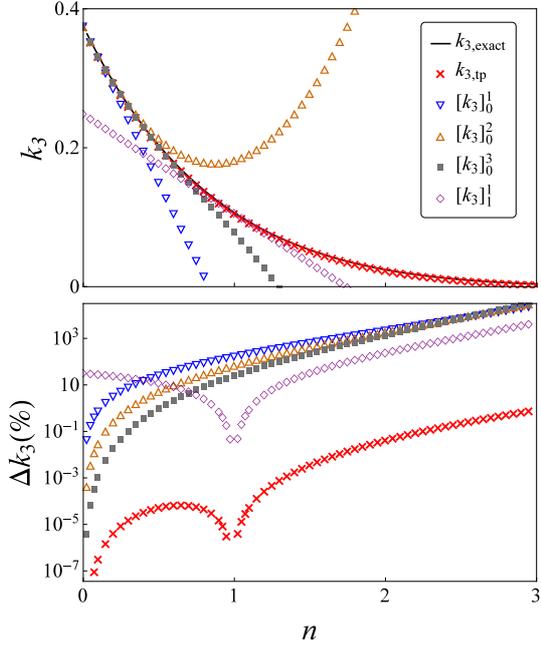}
    \caption{\label{fig:k3_convergence_accuracy}
    In the upper panel, the value of the tidal Love number of degree $l=3$, $k_{3}$, is plotted for
    the numerically exact solution $k_{3,\text{exact}}$ (black line),
    the two-point Pad\'{e} approximant $k_{3,\text{tp}}$ (red cross) in equation (\ref{eqn:klPade}),
    the approximants about $n=0$, $[k_{3}]^{j}_{0}$ in equation (\ref{eqn:polytrope_k3n0})
    for $j=1$ (blue triangle), $j=2$ (orange triangle) and $j=3$ (grey square), and
    the first-order approximation about $n=1$, $[k_{3}]^{1}_{0}$ in equation (\ref{eqn:polytrope_k3n1}) (purple diamond).
    The associated error plot is shown in the lower panel, where
    $\Delta k_{3} = |k_{3} - k_{3,\text{exact}}|/k_{3,\text{exact}} \times 100\%$
    represents the percentage error of the approximant.
    }
\end{figure}

Next, we gauge the accuracy of the perturbation series of $k_{2}$
and $k_{3}$ shown respectively in equations
(\ref{eqn:polytrope_k2n0}), (\ref{eqn:polytrope_k3n0}), 
(\ref{eqn:polytrope_k2n1}), (\ref{eqn:polytrope_k3n1}), as well
as the two-point approximants $k_{2,\text{tp}}$ and
$k_{3,\text{tp}}$. To illustrate the convergence of the
perturbation series of $k_{2}$ and $k_{3}$, we introduce the
symbol $[k_{l}]_{p}^{j}$ to denote the $j$th order partial sum of
the perturbation series of the tidal Love number of degree $l=2$
or $3$ about the perturbation centre $p = 0$ or $1$.  As could be
seen in Figures \ref{fig:k2_convergence_accuracy} and
\ref{fig:k3_convergence_accuracy}, the accuracy improves by two
orders of magnitude for each increment of the perturbation order
near the perturbation centre.  The two-point Pad\'{e} approximants
are formed by merging the third order perturbation results at
$n=0$ and the first order results at $n=1$. The percentage errors
of the two point Pad\'{e} approximants $k_{2,\text{tp}}$ and
$k_{3,\text{tp}}$ are within are $0.39$ per cent and $0.93$ per
cent respectively over the range $n\in[0,3]$.  In particular, the
percentage errors of  $k_{2,\text{tp}}$ and $k_{3,\text{tp}}$
within the range of polytropic index $n\in[0,1]$, between the two
perturbation centres $n=0$ and $n=1$, are $2.5 \times 10^{-5} $
per cent and $7.5 \times 10^{-5} $ per cent respectively (see also
Table~\ref{table:error}).

These accurate approximations of the tidal Love numbers are useful
to astrophysicists in understanding the tidal responses of
polytropic stars, and lay the foundation of an analytical
understanding of the stationarity of the moment-Love relations. We
observe that the tidal Love number decreases monotonically with
increasing $n$ from the incompressible limit. Since the density
tail of polytropes lengthens with increasing polytropic index $n$,
our observation implies that the tidal Love number declines with a
more extended density distribution.

\section{Stationarity of Mutlipolar Moment-Love Relations in the Incompressible Limit} \label{sec:ILoverelations_general}
As reviewed in Section \ref{sec:ILQ_intro},  the mass, the moment
of inertia and the tidal Love number of compact stars (including
both NSs and QSs)  are interconnected in an EOS-independent way to
a high degree of accuracy, commonly referred to as the I-Love
relation \citep{YagiYunes2013,YagiYunes2013Science}. It has also
been observed that the I-Love relation is stationary in the
incompressible limit for several model EOSs such as the modified
Tolman model \citep{Samson_fmode_ILoveQ,Sham_2015} and the
generalised quark matter model \citep{ILoveQ_3,ILoveQ_4}. In this
section, before delving into a detailed study of the I-Love
relation of individual polytropic stars,  we explicitly
demonstrate that the stationarity of I-Love relation in Newtonian
gravity about the incompressible limit in an EOS independent
manner by making use of the perturbative formula
\eqref{eqn:kl_solution_inc} for the tidal
Love number. 
Under our
formulation, the universality is easily generalised to the case
where $l \geq 2$, which is consistent with the results obtained by
\citet{Samson_fmode_ILoveQ}.

\subsection{Moment-Love relation in Newtonian gravity}
As first noted by \citet{YagiYunes2013} for the case of quadrupole
tidal Love number with $l=2$ and later generalised by
\citet{Samson_fmode_ILoveQ} to case with $l \ge 2$, the I-Love
relation, or in general, the multipolar moment-Love relation, in
Newtonian gravity is given by \citep[see equation (4.23)
in][]{Samson_fmode_ILoveQ}:
\begin{equation}\label{multipolar}
  \bar{\lambda}_l = B_l \left(\frac{I_{2l-2}}{M^{2l-1}}\right)^{(2l+1)(2l-2)}
\end{equation}
where the dimensionless multipolar tidal deformability
$\bar{\lambda}_{l}$  is related to the multipolar tidal
deformability $\lambda_{l}$ and the multipolar tidal Love number
$k_l$ by:
\begin{equation}
\bar{\lambda}_l \equiv \frac{\lambda_{l}}{M^{2l+1}} =
\frac{2}{(2l-1)!!} \left(\frac{R}{M}\right)^{2l+1} k_{l},
\end{equation}
and  $I_{k}$ ($k=0,1,2,\ldots$) is essentially  the $k$th mass
moment of a star,
\begin{equation}
I_{k} = \int_{0}^{R} \rho(t) t^{k+2} dt.
\end{equation}
While the form of this multipolar moment-Love relation can be
argued  from (i) dimensional analysis  in Newtonian mechanics
where the speed of light should not appear, and (ii) the intuition
that it should not be affected by the stellar $R$ in an unphysical
way (we will get back to this point in Section
\ref{sec:discussion}), its universality hinges on the observation
that the parameter $B_l$ is almost independent of the EOS as long
as the EOS is sufficiently stiff, i.e., close to the
incompressible limit.

To this end, in the present paper we consider the quantity
\begin{equation} \label{eqn:Uldefinition}
U_{l} \equiv  k_{l}
\left(\frac{\mathcal{M}_{2l-2}}{\mathcal{M}_{0}}\right)^{-(2l+1)/(2l-2)},
\end{equation}
where the dimensionless counterpart of $I_k$, $\mathcal{M}_{k}$,
is defined by:
\begin{equation}\label{eqn:M_k}
\mathcal{M}_{k} = \frac{1}{\rho_{c} R^{k+3}}\int_{0}^{R} \rho(t) t^{k+2} dt = \frac{I_{k}}{\rho_{c} R^{k+3}}.
\end{equation}
It is obvious that the universality of the multipolar moment-Love
relation \eqref{multipolar} is equivalent to the EOS-independency
of $U_{l}$. In geometric units ($G=c=1$), in fact, we have:
\begin{equation}
B_{l} = \frac{2}{(2l-1)!!} (4\pi)^{(2l+1)(2l-2)} U_{l}.
\end{equation}
In terms of the scaled tidal deformability $\bar{\lambda}_{2}$
and the scaled moment of inertia $\bar{I}$, we have:
\begin{equation}
\label{eqn:U2_lambda2_I}
U_{2} = \bigg(\frac{2}{3}\bigg)^{3/2} \bar{\lambda}_{2} \bar{I}^{-5/2}.
\end{equation}
We hint that, in Netwonian gravity, to analyse the universality of
the I-Love-Q relations, it suffices to study the constancy of $U_{2}$
(see Sections \ref{sec:ILQ_intro} and \ref{sec:discussion}).

While such independency has been checked numerically
and analytically for several model EOSs, including polytropic
stars with $n=0$ and $1$ for the case $l=2$ \citep{YagiYunes2013} and
the modified Tolman model  for the case $l \ge 2$
\citep{Samson_fmode_ILoveQ,Sham_2015}, to our knowledge, as yet
there is no generally valid theoretical justification.

\subsection{Stationarity of multipolar moment-Love relations} \label{sec:incompressible_stationarity}
 Our perturbative analysis (\ref{eqn:kl_solution_inc}) of the tidal Love
number from the incompressible limit is applicable to arbitrary
EOSs. We shall use it   to show the stationarity of the
moment-Love relation in an EOS-independent manner.
The variation of $U_{l}$ in the moment-Love relation
\eqref{eqn:Uldefinition} against an arbitrary perturbation
$\varepsilon \rho_1(r)$ in the incompressible limit
is given by the sum of three terms:
\begin{align}
\nonumber & \quad \frac{1}{U_{l}} \frac{d U_{l}}{d
\epsilon}|_{\epsilon =0}
 \equiv \frac{d \ln U_{l}}{d \epsilon}|_{\epsilon =0} \\
& = \left[ \frac{d \ln k_{l}}{d \epsilon} +
\frac{2l+1}{2l-2} \frac{d \ln \mathcal{M}_{0}}{d
\epsilon} - \frac{2l+1}{2l-2} \frac{d \ln
\mathcal{M}_{2l-2}}{d \epsilon}\right]_{\epsilon =0}.
\label{eqn:Ul-var}
\end{align}
From the incompressible limit, the $k$th moment of the density
profile, $\mathcal{M}_{k}$, is related to the $k$th moment of the
density perturbation, $\mu_{k}$, by:
\begin{equation}
\label{eqn:Mk_muk_incompressible}
\mathcal{M}_{k} = \frac{1}{k+3} \bigg[ 1 + \epsilon (k+3) \mu_{k}(R) \bigg].
\end{equation}
By equations (\ref{eqn:kl_solution_inc}) and (\ref{eqn:Mk_muk_incompressible}),
we find that:
\begin{align}
&\label{eqn:klv} \frac{d \ln k_{l}}{d \epsilon}|_{\epsilon =0} = - \frac{3(2l+1)}{2l-2} \mu_{0}(R) + \frac{(2l+1)^2}{2l-2} \mu_{2l-2}(R), \\
& \label{eqn:M0v}\frac{2l+1}{2l-2}\frac{d \ln \mathcal{M}_{0}}{d \epsilon}|_{\epsilon =0} = + \frac{3(2l+1)}{2l-2}\mu_{0}(R), \\
& -\label{eqn:Mlv} \frac{2l+1}{2l-2}\frac{d \ln
\mathcal{M}_{2l-2}}{d \epsilon}|_{\epsilon =0} = -
\frac{(2l+1)^2}{2l-2} \mu_{2l-2}(R),
\end{align}
Hence, we conclude that $U_{l}$ is stationary against an
arbitrary perturbation $\varepsilon \rho_1(r)$ in the
incompressible limit:
\begin{equation}
\label{eqn:Ul_incompressible_stationarity}
\frac{d U_{l}}{d \epsilon}|_{\epsilon =0} = 0,
\end{equation}
and at the incompressible limit:
\begin{equation}
\label{eqn:Ul_0thorder} U_{l}(\epsilon = 0) = \frac{3}{4(l-1)}
\bigg( \frac{2l+1}{3} \bigg)^{(2l+1)/(2l-2)}.
\end{equation}

We stress that the stationarity of the multipolar moment-Love
relation in the incompressible limit is general, because it does
not depend specifically on the functional form of the density
perturbation $\rho_{1}$.  Hence, the multipolar moment-Love
relation is insensitive to the equation of state near the
incompressible limit. The stationarity of $U_{l}$ and the
universality of the multipolar moment-Love relation are
remarkable, in the sense that two seemingly unrelated physical
quantities, the tidal Love number and the mass moments, are
related in an EOS-independent manner from the incompressible
limit. Intuitively speaking, to the first order, the tidal Love
numbers $k_{l}$ respond to the density variation $\rho_1(r)$
through its zeroth moment $\mu_{0}(R)$ and the $(2l-2)$th moment
$\mu_{2l-2}(R)$. While the former is purely a mass effect due to
the density variation, the latter further measures the spatial
extent of $\rho_1(r)$. However, these two changes are exactly
cancelled by the corresponding responses in the moments
$\mathcal{M}_{0}$ and $\mathcal{M}_{2l-2}$, which both appear in
the multipolar moment-Love relations. This observation is indeed
an independent corroboration of the correctness of the form of
such EOS-insensitive relations.

\section{Moment-Love relations of polytropic stars} \label{sec:ILoverelations_polytropes}
The perturbative formulae for the tidal Love numbers $k_{2}$ and
$k_{3}$ of polytropes obtained  in Section
\ref{sec:polytrope_perturbation} allow us to examine the I-Love
relation (or the multipolar moment-Love relation) of polytropic
stars analytically. The analytical results, which are  also
consistent with the numerical ones, show that $U_2$ in the I-Love
relation is, within $0.4$ per cent, a constant near the incompressible
limit where $n\in[0,1]$. The deviation grows sharply to $6$ per cent when
$n=2$, and reaches $25 $ per cent at $n=3$ (see
Table~\ref{table:error}). We shall show that the constancy of
$U_2$ near the incompressible limit where $n\in[0,1]$ is actually
attributable to the universal stationary point at $n=0$ and, in
addition, the presence of a secondary stationary point at $n
\approx 0.4444$. Numerical and analytical results of the
moment-Love relation for $l=3$ are also presented.  Compared to
the I-Love relation with $l=2$, the value of $U_3$ appearing in
the moment-Love relation for $l=3$ varies more significantly than
that of $U_2$. While $U_3$ is stationary at $n=0$, there is no
secondary stationary point for $n\in[0,1]$.

\subsection{I-Love relation of polytropic stars}
We first focus on the I-Love relation of polytropic stars, where
$l=2$, and
 $U_{2} = k_{2} (\mathcal{M}_{2}/ \mathcal{M}_{0})^{-5/2}$.
As an accurate analytical approximation to the tidal Love number
$k_{2}$, the  two-point Pad\'{e} approximant
in equation (\ref{eqn:klPade}),
has been derived, it remains to determine the moment ratio
$\mathcal{M}_{2}/\mathcal{M}_{0}$. Physically, $\mathcal{M}_{0}$
and $\mathcal{M}_{2}$ are proportional to the mass $M$ and the
moment of inertia $I$ respectively, by $M = 4 \pi \rho_{c} R^{3}
\mathcal{M}_{0}$ and $I = (8 \pi/3) \rho_{c} R^{5}
\mathcal{M}_{2}$.  Hence, the fraction of the dimensionless
moments, $\mathcal{M}_{2}/\mathcal{M}_{0}$, is equal to $3 I /(2 M
R^2)$.

In fact, we find that the perturbation series of the fraction
$\mathcal{M}_{2l-2} / \mathcal{M}_{0}$ is more accurate than
individual moments because for each singularity in
$\mathcal{M}_{0}$ inherited from the density profile, there is a
singularity of the same kind in $\mathcal{M}_{2l-2}$. Following
directly from the solution of the LEE with $n=0$, the leading
order approximation of $\mathcal{M}_{k}(n)$ from the
incompressible limit is given by:
\begin{align}
\nonumber \mathcal{M}_{k}(n) & = \frac{1}{\pi^3} \int_{0}^{\pi} (1 - \frac{z^2}{\pi^2} )^{n} z^{k+2} dz  + O[n^2] \\
& = \frac{\Gamma \left((k+3)/2\right) \Gamma (1+n)}{2 \Gamma
\left( (k + 2 n + 5)/2 \right)} + O[n^2],
\end{align}
for  $n > -1$. As could be seen, the $k$th moment contains simple 
poles at $n = -1, -2, -3, \ldots$ due to the term $\Gamma(1+n)$.
Such singularities are caused by the density profile $(1 -
z^2/\pi^2)^{n}$, which is integrable over the domain of interest
for $n > -1$.  As both the zeroth moment $\mathcal{M}_{0}$ and the
$(2l-2)$th moment $\mathcal{M}_{2l-2}$ are directly proportional
to the density profile, such singularities due to the
$\Gamma(1+n)$ cancel out in the fraction $\mathcal{M}_{2l-2} /
\mathcal{M}_{0}$. While the $n$-series approximation of
$\mathcal{M}_{0}$ and $\mathcal{M}_{2}$ have a
common radius of convergence of $1$ due to the nearest singularity
at $n=-1$, the fractions $\mathcal{M}_{2}/\mathcal{M}_{0}$ has
a larger interval of validity.

Therefore, it is more advantageous for us to determine the
perturbation series for the fractions
$\mathcal{M}_{2}/\mathcal{M}_{0}$, instead of the moments
themselves, about $n=0$ and $n=1$, respectively, and then derive a
two-point Pad\'{e} approximant to connect these series. Using the
results of SDEM at $n=0$ and $n=1$ and going through some standard
but a bit tedious calculations, we arrive at the following
results: (i) an expansion of $\mathcal{M}_{2}/\mathcal{M}_{0}$
about $n=0$,
\begin{align}
\nonumber \frac{\mathcal{M}_{2}}{\mathcal{M}_{0}} & = \frac{3}{5}-\frac{6 n}{25}+ \bigg(\frac{1496}{375}-\frac{2 \pi ^2}{5} \bigg) n^2 \\
\nonumber & \quad + \bigg[ -\frac{63504}{625} + \frac{202 \pi ^2}{75} + \frac{312}{5}\zeta(3) \bigg] n^3 + O[n^4] \\
\label{eqn:M0M2n0} & \approx 3/5 - 6n/25 + 0.041491573 n^2 -
0.015914723 n^3 + O[n^4],
\end{align}
(ii) an expansion of $\mathcal{M}_{2}/\mathcal{M}_{0}$ about $n=1$
\begin{align}
\nonumber \frac{\mathcal{M}_{2}}{\mathcal{M}_{0}} &
=1-\frac{6}{\pi ^2} - \frac{3 (n-1)}{\pi ^3} [-5 \pi +4 \pi  \ln 2 + 2 \text{Si}(\pi)+\text{Si}(2 \pi )]\\ \nonumber &\quad+O[(n-1)^2]\\
\label{eqn:M0M2n1} & \approx 0.39207290 - 0.18147228 (n - 1) +
O[(n-1)^2],
\end{align}
and (iii) a two-point Pad\'{e} approximant obtained by merging the
results shown in equations (\ref{eqn:M0M2n0})  and
(\ref{eqn:M0M2n1}),
\begin{equation}
\label{eqn:moment_ratio_tp}
[\frac{\mathcal{M}_{2l-2}}{\mathcal{M}_{0}}]_{\text{tp}} = (5-n)^3
\frac{b_{1} + n b_{2} + n^2 b_{3}  + n^3 b_{4} }{1000 + n b_{5} +
n^2 b_{6}},
\end{equation}
where $\text{Si}(x)$ is the integral of sine function \citep[see,
e.g.,][and equation \eqref{eq:Si}]{Math_handbook}, and the
constants $b_{1}, b_{2}, \ldots, b_{6}$ for the case $l=2$ are
determined from equations (\ref{eqn:M0M2n0}) and
(\ref{eqn:M0M2n1}) and tabulated in Table
\ref{table:2}.  In equation
\eqref{eqn:moment_ratio_tp} we have incorporated an empirical
factor of $(5-n)^3$, which is found to greatly enhance the
accuracy of the two-point approximant. The motivation for the form
of such an empirical factor can be found in Appendix \ref{sec:n5}.
\begin{table}
\centering \caption{\label{table:2} The numerical values of the
constants $b_{1}, b_{2}, \ldots, b_{6}$ appearing in
\eqref{eqn:moment_ratio_tp}.}
 \begin{tabular}{|l |l | l |}
    \hline
            & $l=2$                 & $l=3$ \\
    \hline
    $b_{1}$ & $24/5$                & $24/7$                 \\
    $b_{2}$ & $3.452950061777$      & $1.605541075759$      \\
    $b_{3}$ & $0.08417922198458$    & $0.5058524906146$      \\
    $b_{4}$ & $0.01816712663676$    & $0.02106526231457$      \\
    $b_{5}$ & $519.3645962035$      & $553.9970994773$     \\
    $b_{6}$ & $-155.4882028232$     & $-137.1355972892$     \\
    \hline
 \end{tabular}
\end{table}

\begin{figure}
    \includegraphics[width = 0.45 \textwidth]{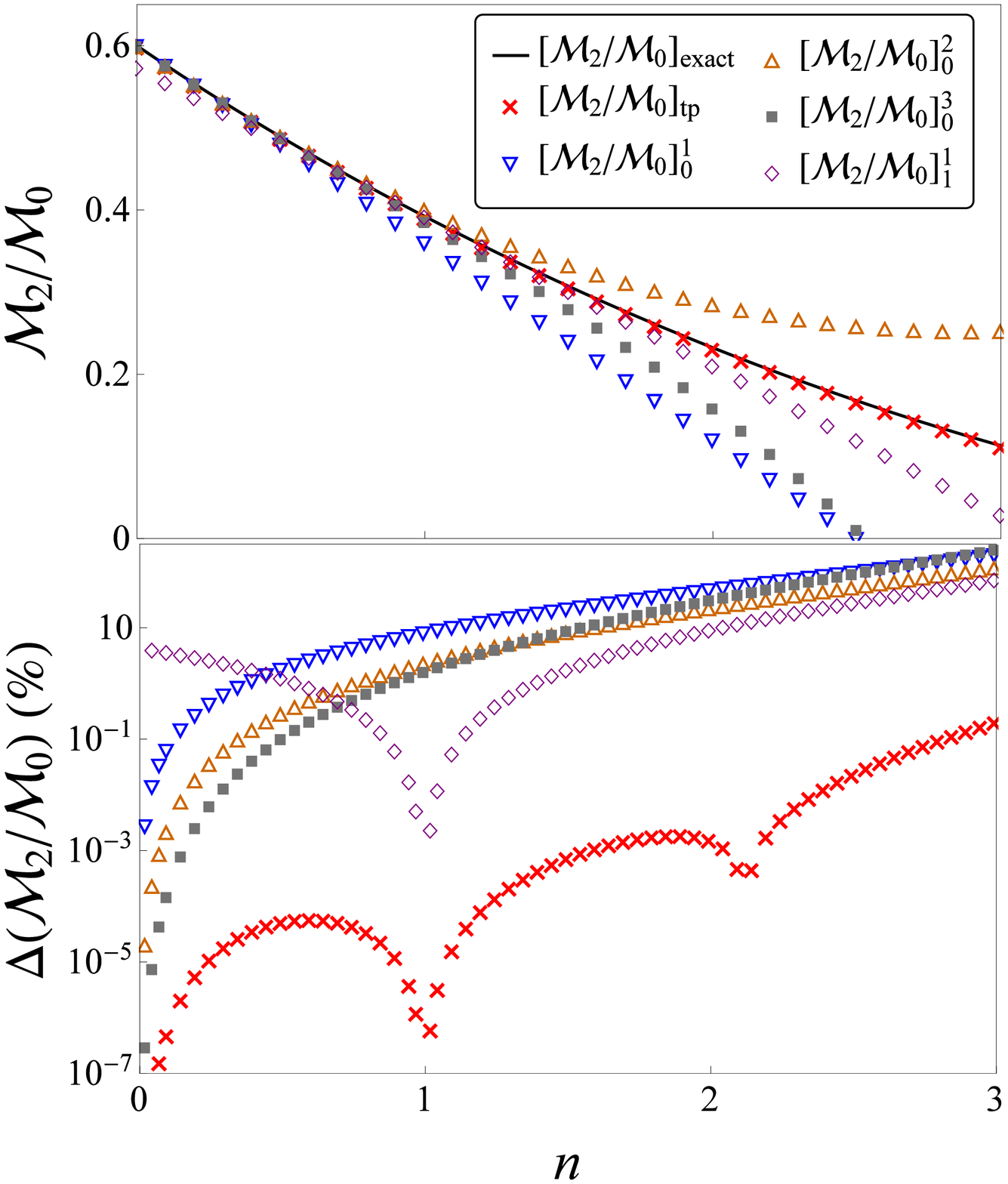}
    \caption{\label{fig:M2oM0}
    In the upper panel, the value of the moment fraction, $\mathcal{M}_{2}/\mathcal{M}_{0}$, is plotted for
    the numerically exact solution $[\mathcal{M}_{2}/\mathcal{M}_{0}]_{\text{exact}}$ (black line),
    the two-point Pad\'{e} approximant $[\mathcal{M}_{2}/\mathcal{M}_{0}]_{\text{tp}}$ (red cross) in equation (\ref{eqn:moment_ratio_tp}),
    the approximants about $n=0$, $[\mathcal{M}_{2}/\mathcal{M}_{0}]^{j}_{0}$ in equation (\ref{eqn:M0M2n0})
    for $j=1$ (blue triangle), $j=2$ (orange triangle) and $j=3$ (grey square), and
    the first-order approximation about $n=1$, $[\mathcal{M}_{2}/\mathcal{M}_{0}]^{1}_{1}$ in equation (\ref{eqn:M0M2n1}) (purple diamond).
    The associated error plot is shown in the lower panel, where
    $\Delta (\mathcal{M}_{2}/\mathcal{M}_{0}) = |[\mathcal{M}_{2}/\mathcal{M}_{0}] / [\mathcal{M}_{2}/\mathcal{M}_{0}]_{\text{exact}} - 1| \times 100\%$
    represents the percentage error of the approximant.
    }
\end{figure}

We gauge the accuracy of the perturbation series of
$\mathcal{M}_{2}/\mathcal{M}_{0}$ shown respectively in equations
(\ref{eqn:M0M2n0}) and (\ref{eqn:M0M2n1}), as well as the
two-point approximant
$[\mathcal{M}_{2}/\mathcal{M}_{0}]_{\text{tp}}$ in equation
(\ref{eqn:moment_ratio_tp}). To illustrate the convergence of the
perturbation series of $\mathcal{M}_{2}/\mathcal{M}_{0}$, we
introduce the symbol $[\mathcal{M}_{2}/\mathcal{M}_{0}]_{p}^{j}$
to denote the $j$th order partial sum of the perturbation series
of the the moment fraction about $p = 0$ or $1$.  As could be seen
in Figure \ref{fig:M2oM0}, the accuracy improves by an order of
magnitude for each increment of the perturbation order near the
perturbation centre.  
The percentage errors of the
two point Pad\'{e} approximants
$[\mathcal{M}_{2}/\mathcal{M}_{0}]_{\text{tp}}$ are within are
$0.21$ per cent over the range $n\in[0,3]$.  In particular, the
percentage errors of
$[\mathcal{M}_{2}/\mathcal{M}_{0}]_{\text{tp}}$ within the range
of polytropic index $n\in[0,1]$, between the two perturbation
centres $n=0$ and $n=1$, is just $6.1 \times 10^{-5}$ per cent.

We define the analytical two-point approximant of $U_{2}$ by
$U_{2,\text{tp}} = k_{2, \text{tp}} [\mathcal{M}_{2}
/\mathcal{M}_{0}]_{\text{tp}}^{-5/2}$ and first analyse the I-Love
relation near the incompressible limit for $n\in[0,1]$.
\begin{figure}
    \includegraphics[width = 0.45 \textwidth]{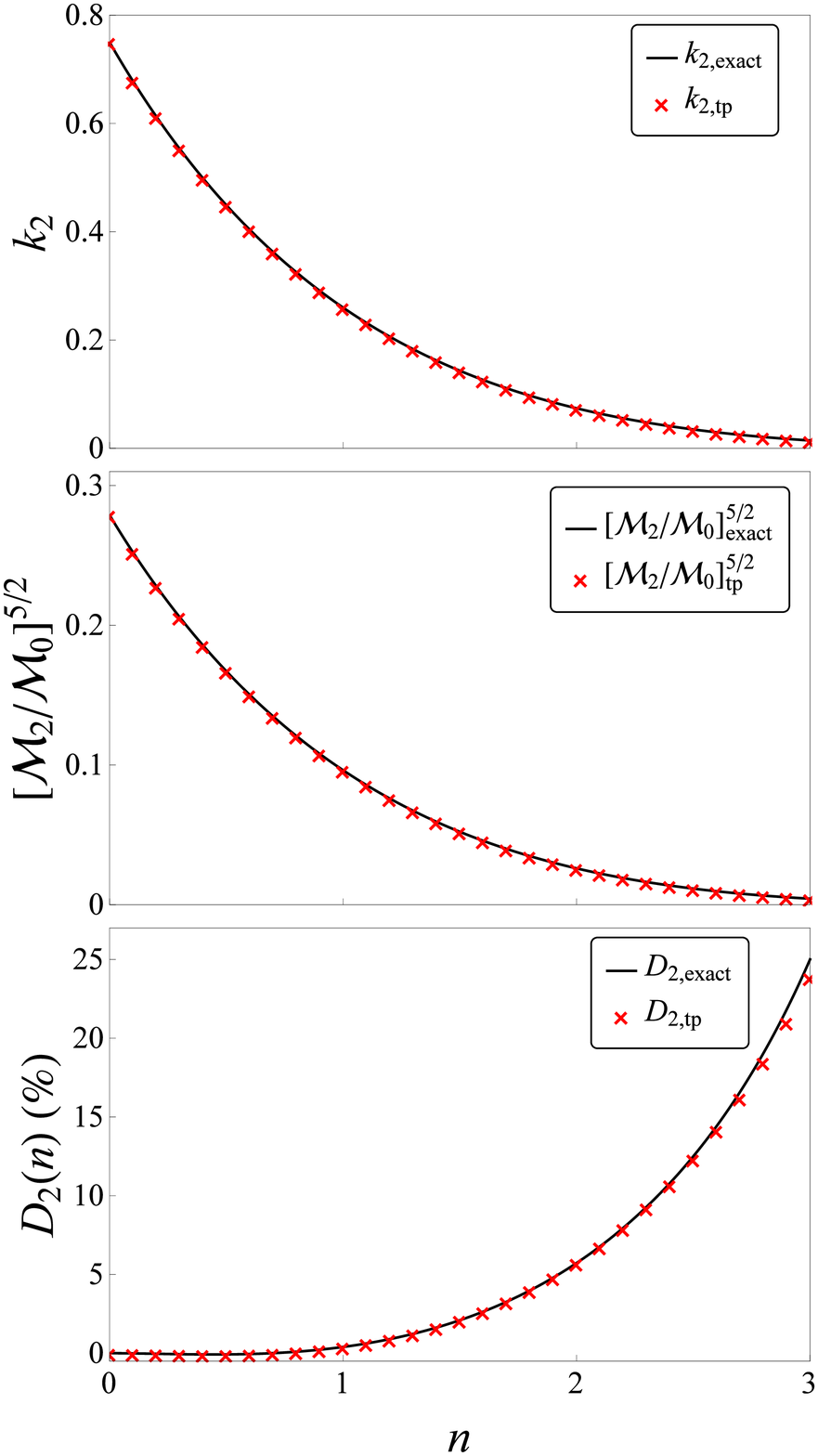}
    \caption{\label{fig:D2_quotient_explanation}
    In the upper panel, the quadrupole tidal Love number $k_{2}$ is plotted over the range of stable polytropes for $n\in[0,3]$,
    for the numerically exact solution $k_{2, \text{exact}}$ (black line), and
    the two-point Pad\'{e} approximant $k_{2, \text{tp}}$ (red cross) in equation (\ref{eqn:klPade}).
    In the middle panel, the moment fraction raised to the power $5/2$, $[\mathcal{M}_{2} / \mathcal{M}_{0}]^{5/2}$ is plotted over the same range for $n\in[0,3]$,
    for the numerically exact solution $[\mathcal{M}_{2} / \mathcal{M}_{0}]^{5/2}_{\text{exact}}$ (black line), and
    the two-point Pad\'{e} approximant $[\mathcal{M}_{2} / \mathcal{M}_{0}]^{5/2}_{\text{tp}}$ (red cross) in equation (\ref{eqn:moment_ratio_tp}).
    In the lower panel, the percentage variation of the I-Love relation $U_{2}$ over the range of stable polytropes for $n\in[0,3]$ is plotted for
    the numerically exact solution $D_{2,\text{exact}} = [U_{2,\text{exact}}/ U_{2}(0) - 1] \times 100 \% $ (black line) and
    the two-point approximant $D_{2,\text{tp}} = [U_{2,\text{tp}}/ U_{2}(0) - 1] \times 100 \% $ (red cross).
    }
\end{figure}
In Figure \ref{fig:D2_quotient_explanation}, the numerical results
and the analytical calculations consistently show that the I-Love
relation $U_{2}$ is remarkably flat near the incompressible limit
for $n\in[0,1]$, and the variation is within $0.4$ per cent. In addition,
the constancy of the I-Love relation is indeed attributable to the
existence of multiple stationary points for $n\in[0,1]$.
Numerically, $U_{2}$ has stationary points at $n=0$ and $n \approx
0.4444$.  Our analytical calculations can capture such features as
well. Demanding $dU_{2,\text{tp}}/dn = 0$, we find stationary
points $n=0$ and $n \approx 0.444364$, which agrees nicely with
the numerical value $n \approx 0.4444$.

Alternatively, we can use equations \eqref{eqn:polytrope_k2n0} and
\eqref{eqn:M0M2n0} to find an expansion for $U_{2}$ about the
incompressible limit with $n=0$:
\begin{align}
\nonumber U_{2}(n) & = \sqrt{\frac{5}{3}} \bigg\{ \frac{25}{12} + n^2 \bigg( \frac{25 \pi ^2}{36}-\frac{1487}{216} \bigg) + n^3 \bigg[ \frac{59093}{300} \\
\nonumber & \quad - \frac{599 \pi^2}{108} - \frac{325}{3} \zeta (3) + \big(-\frac{1087}{27} + \frac{10 \pi^2}{9}\big) \ln 2 \\
\nonumber & \quad + 24 \ln ^2 2 -\frac{20}{3} \ln^3 2 - \frac{375}{8} J(2,2;4)+\frac{1125}{4} J(2,4;4)  \\
\nonumber & \quad - \frac{1875}{8} J(4,4;4) \bigg] \bigg\} \\
\label{eqn:U2n0} & \approx 2.6895718 - 0.039204033 n^2 +
0.084829962 n^3 + O[n^4],
\end{align}
which is normalised by the value at the incompressible limit
$n=0$:
\begin{equation}\label{eqn:U2n0_norm}
\frac{U_{2}(n)}{U_{2}(n=0)} = 1 - 0.014576311 n^2 + 0.031540323
n^3 + O[n^4].
\end{equation}
The first order variation with respect to $n$ vanishes at $n=0$ as
expected. Moreover, the existence of another stationary point at
$n\approx0.3081$ close to the first one where $n=0$ is
well-predicted by the third order perturbation result shown in
equation \eqref{eqn:U2n0} or \eqref{eqn:U2n0_norm}.  We note that
the coefficients of the $n^2$ and $n^3$ terms in the Taylor series
expansion of $U_{2}(n)/U_{2}(n=0)$ are much smaller than unity,
and the second order term has an opposite sign from the zeroth and
the third order terms. These, again, hint at an extended interval
of validity of the constancy of the I-Love relation from the
incompressible limit.

In Figure \ref{fig:D2_quotient_explanation}, from $n=0$ to $n=1$,
while both the quadrupole Love number $k_{2}$ and the moment
fraction $[\mathcal{M}_{2}/\mathcal{M}_{0}]^{5/2}$ roughly drop by
a factor of $3$, the value of $U_{2}(n)$ defined in the I-Love
relation is remarkably stable with the variation being within
$0.4$ per cent. In addition, over the extended range of stable polytropes
from $n=0$ to $n=3$, while the quadrupole Love number $k_{2}$ and
the moment fraction $[\mathcal{M}_{2}/\mathcal{M}_{0}]^{5/2}$
simultaneously decrease by two orders of magnitude, $U_{2}(n)$
only changes by $25 $ per cent.

Next, we analyse the breakdown of I-Love relation. In Figure
\ref{fig:D2_quotient_explanation}, both the numerical and the
analytical results show that the $U_{2}(n)$ in the I-Love relation
significantly deviates from its incompressible counterpart,
$U_{2}(0)$, for polytropes of larger polytropic indices. As a
matter of fact, $D_2 \equiv [U_{2}(n)/U_{2}(0)-1)] \times 100 \%$
grows from $0.4$ per cent via $6.0$ per cent to $25$ per cent, when $n$ increases from
$1$ via $2$ to $3$.  It illustrates the breakdown of universality
when the EOS is soft, or the density distribution is more
extended.

It is remarkable that, as  shown in Figure
\ref{fig:D2_quotient_explanation}, the two-point approximants
derived in the present paper can accurately capture the variation
of the quadrupole tidal Love number and the scaled moment
$[\mathcal{M}_{2}/\mathcal{M}_{0}]$. Through these two
approximants, we can demonstrate the validity of the I-Love
relation as well as its limitation, which clearly pinpoints  the
physical origin of such universality.

\subsection{Octupolar moment-Love relation of polytropic stars}
Next, we repeat the calculations for the moment-Love relation for
$l=3$.  From equation (\ref{eqn:Uldefinition}), we see that $U_{3}
= k_{3} (\mathcal{M}_{4}/\mathcal{M}_{0})^{-7/4}$.  The
approximation of $k_{3}$ has been determined in Section
\ref{sec:polytrope_perturbation}. To analytically analyse the the
moment-Love relation $U_{3}$, it remains to evaluate the moment
fraction $\mathcal{M}_{4}/\mathcal{M}_{0}$.  We determine the
perturbation expansion of $\mathcal{M}_{4}/\mathcal{M}_{0}$ about
the incompressible limit $n=0$:
\begin{align}\label{eqn:M0M4n0}
\frac{\mathcal{M}_{4}}{\mathcal{M}_{0}} & = \frac{3}{7}-\frac{72 n}{245}+ n^2\bigg(\frac{147382}{25725}-\frac{4 \pi ^2}{7}\bigg) \nonumber \\
& \quad + n^3 \bigg(-\frac{432846317}{2701125}+\frac{908 \pi ^2}{245}+\frac{720 \zeta (3)}{7} \bigg) + O[n^4] \nonumber \\
& \approx  \frac{3}{7} + \frac{72n}{245} + 0.089361139 n^2 -
0.028580246 n^3+ O[n^4],
\end{align}
as well as the case  $n=1$,
\begin{align}
\label{eqn:M0M4n1}
\frac{\mathcal{M}_{4}}{\mathcal{M}_{0}} & =
1+\frac{120}{\pi ^4}-\frac{20}{\pi ^2} + \frac{10(n-1)}{\pi^5} \bigg[94 \pi -5 \pi ^3    \nonumber \\
& \quad  - 4 \pi(12 - \pi^2 ) \ln 2 - 24 \text{Si}(\pi )-
(12+\pi^2) \text{Si}(2 \pi ) \nonumber \\
& \quad - 21 \pi \zeta (3) \bigg] +O[(n-1)^2]\nonumber \\& \approx 0.20549420-0.16501233(n-1) + O[(n-1)^2],
\end{align}
and merge these two expansions together to obtain a two-point
Pad\'{e} approximant,
$[\mathcal{M}_{4}/\mathcal{M}_{0}]_{\text{tp}}$, whose explicit
form
 is given in
equation (\ref{eqn:moment_ratio_tp}) with the constants $b_{1},
b_{2}, \ldots , b_{6}$ listed in
Table~\ref{table:2}. Then, a two-point approximant
for the $l=3$ moment-Love relation can be formed, namely, $U_{3,
\text{tp}} = k_{3, \text{tp}} [\mathcal{M}_{4} /
\mathcal{M}_{0}]_{\text{tp}}^{-7/4}$.

\begin{figure}
    \includegraphics[width = 0.45 \textwidth]{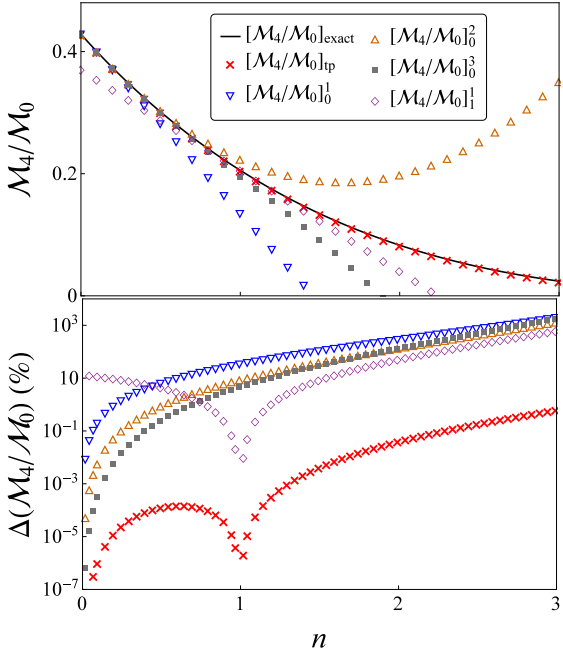}
    \caption{\label{fig:M4oM0}
    In the upper panel, the value of the moment fraction, $\mathcal{M}_{4}/\mathcal{M}_{0}$, is plotted for
    the numerically exact solution $[\mathcal{M}_{4}/\mathcal{M}_{0}]_{\text{exact}}$ (black line),
    the two-point Pad\'{e} approximant $[\mathcal{M}_{4}/\mathcal{M}_{0}]_{\text{tp}}$ (red cross) in equation (\ref{eqn:moment_ratio_tp}),
    the approximants about $n=0$, $[\mathcal{M}_{4}/\mathcal{M}_{0}]^{j}_{0}$ in equation (\ref{eqn:M0M4n0})
    for $j=1$ (blue triangle), $j=2$ (orange triangle) and $j=3$ (grey square), and
    the first-order approximation about $n=1$, $[\mathcal{M}_{4}/\mathcal{M}_{0}]^{1}_{1}$ in equation (\ref{eqn:M0M2n1}) (purple diamond).
    The associated error plot is shown in the lower panel, where
    $\Delta (\mathcal{M}_{4}/\mathcal{M}_{0}) = |(\mathcal{M}_{4}/\mathcal{M}_{0})/ (\mathcal{M}_{4}/\mathcal{M}_{0})_{\text{exact}} - 1| \times 100\%$
    represents the percentage error of the approximant.
    }
\end{figure}

The accuracy of the perturbation series in equations
(\ref{eqn:M0M4n0}) and (\ref{eqn:M0M4n1}), and the two-point
approximant in equation (\ref{eqn:moment_ratio_tp}) is studied in
Figure~\ref{fig:M4oM0}. As introduced previously in the case with $l=2$, 
the symbol
$[\mathcal{M}_{4}/\mathcal{M}_{0}]_{p}^{j}$ 
denotes the $j$th order partial sum of the perturbation series of
the the moment fraction about $p = 0$ or $1$. 
Figure \ref{fig:M4oM0} clearly shows that the accuracy of
$[\mathcal{M}_{4}/\mathcal{M}_{0}]$ improves by an order of
magnitude for each increment of the perturbation order near the
perturbation centre $p = 0$ or $p=1$.  
As a matter of fact, the percentage error of
$[\mathcal{M}_{4}/\mathcal{M}_{0}]_{\text{tp}}$ 
between the two perturbation centres $n=0$ and $n=1$  is less than
 $1.6 \times 10^{-4}$ per cent. Even throughout the entire
physical range $n\in[0,3]$, the percentage error of such a
two-point Pad\'{e} approximant
is still bounded by $0.65$ per cent, which is impressive.

\begin{figure}
    \includegraphics[width = 0.45 \textwidth]{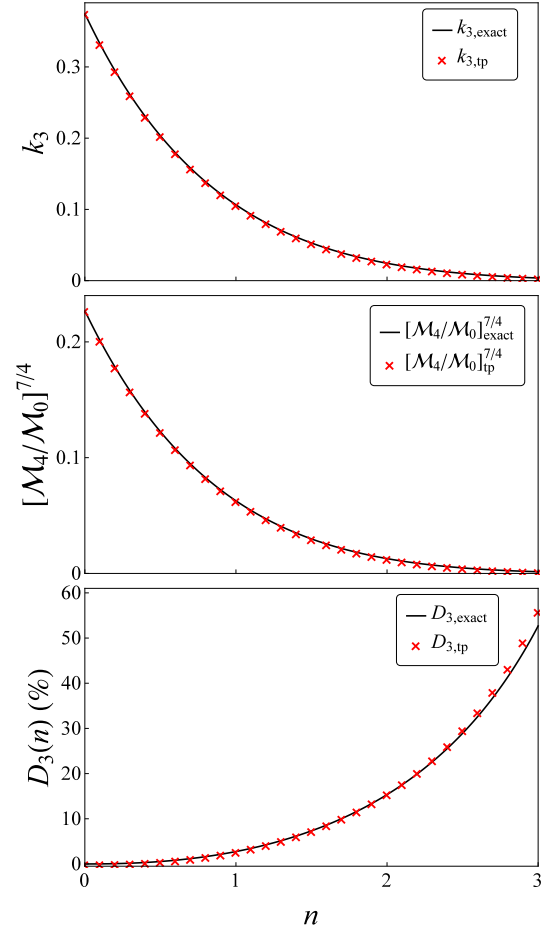}
    \caption{\label{fig:D3_quotient_explanation}
    In the upper panel, the quadrupole tidal Love number $k_{3}$ is plotted over the range of stable polytropes for $n\in[0,3]$,
    for the numerically exact solution $k_{3, \text{exact}}$ (black line), and
    the two-point Pad\'{e} approximant $k_{3, \text{tp}}$ (red cross) in equation (\ref{eqn:klPade}).
    In the middle panel, the moment fraction raised to the power $7/4$, $[\mathcal{M}_{4} / \mathcal{M}_{0}]^{7/4}$ is plotted over the same range for $n\in[0,3]$,
    for the numerically exact solution $[\mathcal{M}_{4} / \mathcal{M}_{0}]^{7/4}_{\text{exact}}$ (black line), and
    the two-point Pad\'{e} approximant $[\mathcal{M}_{4} / \mathcal{M}_{0}]^{7/4}_{\text{tp}}$ (red cross) in equation (\ref{eqn:moment_ratio_tp}).
    In the lower panel, the percentage variation of the third moment-Love relation $U_{3}$ over the range of stable polytropes for $n\in[0,3]$ is plotted for
    the numerically exact solution $D_{3,\text{exact}} = [U_{3,\text{exact}}/ U_{3}(0) - 1] \times 100 \% $ (black line) and
    the two-point approximant $D_{3,\text{tp}} = [U_{3,\text{tp}}/ U_{3}(0) - 1] \times 100 \% $ (red cross).
    }
\end{figure}

In parallel to the $l=2$ case, we proceed to study the moment-Love
relation for the $l=3$ case with the  two-point two-point Pad\'{e}
approximants for the tidal Love number and the moment ratio. We
first consider the relation in the range near the incompressible
limit for $n \in [0,1]$, and then extend our investigation to the
entire range of stable polytropes for $n \in [0,3]$. In Figure
\ref{fig:D3_quotient_explanation}, the numerical and the
analytical results consistently show that the variation of $U_{3}$
is within $2.8$ per cent near the incompressible limit for $n \in [0,1]$.
However, over the extended range of stable polytropes from the
incompressible limit at $n=0$ to the verge of stability at $n=3$, while
both the Love number of degree $l=3$, $k_{3}$, and the moment
fraction, $[\mathcal{M}_{4}/\mathcal{M}_{0}]$, decrease by two
orders of magnitude, the maximum change of $U_3(n)$ the
moment-Love relation is relatively mild at $53 $ per cent. Compared to
$U_{2}$ in the I-Love relation , the variation of $U_{3}$ is more
significant. Besides,
 there is only one stationary point located at $n=0$, where 
the stationary point at $n=0$ is
expected analytically by our analysis in Section
\ref{sec:incompressible_stationarity}, throughout the whole
physical range where $n \in [0,3]$. To better understand the
absence of the secondary stationary point, we can find the Taylor
series of $U_{3}(n)$ about the incompressible limit $n=0$ up to
the third order in $n^3$ from equations \eqref{eqn:polytrope_k3n0}
and \eqref{eqn:M0M4n0}:
\begin{align}
\nonumber U_{3}(n) & = \left(\frac{7}{3}\right)^{3/4} \bigg\{ \frac{7}{8} + n^2 \left(\frac{7 \pi ^2}{24}-\frac{72269}{25200}\right) + n^3 \bigg[ \frac{3208660909}{37044000} \\
\nonumber & \quad -\frac{19 \pi ^2}{28} + \left(-\frac{289253}{11025}+\frac{2 \pi ^2}{3}\right) \ln  2 + \frac{536}{35} \ln ^2 2 \\
\nonumber & \quad - 4 \ln^3 2 - \frac{119 \zeta (3)}{2} + \frac{441}{32} J(2,2;6) + \frac{1323}{16} J(2,6;6) \\
\nonumber & \quad  - \frac{3087}{32} J(6,6;6) \bigg] \bigg\} + O[n^4] \\
\label{eqn:U3n0} & \approx 1.6519262 + 0.020421879 n^2 +
0.043179362 n^3 + O[n^4],
\end{align}
which is normalised by the value at the incompressible limit
$n=0$:
\begin{equation}
\frac{U_{3}(n)}{U_{3}(n=0)} = 1 + 0.012362465 n^2 + 0.026138796
n^3 + O[n^4].
\end{equation}
Using the third order perturbation of $U_{3}(n)$ shown above and
setting $dU_{3}/dn = 0$, we obtain two solutions $n = 0$ and $n =
-0.3153$. The latter solution lies outside the physical domain of
interest, and our analytical results thus consistently predict the
absence of a secondary stationary point.  In stark contrast to the
case where $l=2$, the second and the third order terms carry the
same sign as the zeroth order term, suggesting  a more pronounced
variation around the incompressible limit and the absence of a
secondary stationary point.

Next, we study the octupole moment-Love relation over an extended
range of polytropic index for $n\in[0,3]$.  In Figure
\ref{fig:D3_quotient_explanation}, both the numerical and the
analytical results show that $U_3(n)$ in such a relation varies
significantly for polytropes of larger polytropic indices.  In
particular, our numerical results show that the percentage change
of $U_3(n)$, $D_{3,\text{exact}} = [U_{3,\text{exact}}/ U_{3}(0) -
1] \times 100 \% $, 
are $2.8$ per cent, $15.4$ per cent and $52.7$ per cent at $n=1$,
$n=2$ and $n=3$, respectively. We, again, observe the breakdown of
the universality for soft polytropic stars.

\section{Discussion and Conclusion} \label{sec:discussion}
The I-Love relation has been a hot topic of scholarly account to
date \citep[see][for a review on this topic]{Yagi:2016bkt}.
Various compact stellar objects, including both NSs and QSs,
appear to follow the same relation between the moment of inertia
and the tidal Love number. Its physical origin is certainly an
interesting issue. It has been observed and proved in the cases of
the modified Tolman model and self-bound stars, which both possess
non-zero surface mass density, that the I-Love relation is
stationary in the incompressible limit
\citep{Samson_fmode_ILoveQ,Sham_2015,ILoveQ_3,ILoveQ_4}. Along a
similar line of thought, using our perturbation solutions to the
tidal Love numbers and the mass moments, we have established in
the present paper that the I-Love relation is stationary with
respect to arbitrary density changes in the incompressible limit,
and generalised the stationarity to cases where $l\geq2$ (see
Section~\ref{sec:incompressible_stationarity}).  Our proof is
generally valid in the sense that there is no need to confine the
variation to a certain class (e.g., from a modified Tolman model
(self-bound) star to another modified Tolman model (self-bound)
star as in the previous studies mentioned above. In particular, we
have pinpointed the physical nature of such an EOS-independent
stationarity of the I-Love relation, which is attributable to the
cancellation of the moments of density variation to the first
order in the incompressible limit.

In Newtonian gravity, to study the universality of the I-Love-Q
relation, it suffices to understand the constancy of $U_{2} =
k_{2} [ \mathcal{M}_{2} / \mathcal{M}_{0} ]^{-5/2}$. By the
relation $\bar{\lambda}_{2} = \bar{Q} \bar{I}^2$ and equation
(\ref{eqn:U2_lambda2_I}), we obtain the following constitutive
relations, in terms of the scaled quadrupole moment $\bar{Q}$ and
the scaled moment of inertia $\bar{I}$:
\begin{equation}
U_{2} = \bigg(\frac{2}{3}\bigg)^{3/2} \bar{Q} \bar{I}^{-1/2},
\end{equation}
and in terms of the scaled tidal deformability $\bar{\lambda}_{2}$
and the scaled quadrupole moment $\bar{Q}$:
\begin{equation}
U_{2} = \bigg(\frac{2}{3}\bigg)^{3/2} \bar{\lambda}_{2}^{-1/4} \bar{Q}^{+5/4}.
\end{equation}
Therefore, the EOS-insensitivity  of $U_2$, the I-Love relation
studied in depth in the present paper, readily leads to the
universality of the I-Love-Q relations.

One of the major breakthroughs achieved in the present paper is
the establishment of a perturbation expansion for the  tidal Love
number, which is divergence-free and enables us to evaluate the
tidal Love number of an arbitrary star with zero/non-zero surface
density from that of an incompressible star. Through balancing the
singularities of the tidal field and density profile, we have
proposed the modified potential $h(r) = H(r) / m(r)$.  The
governing equations of the modified potential and its associated
logarithmic derivative,  (\ref{eqn:h}) and (\ref{eqn:y}), are
singularity-free, upon which we have developed the perturbation
expansion. As a result, we can express the perturbative
corrections to the tidal Love number in terms of the moments and
the overlap integrals of the density variation (see equation
(\ref{eqn:kl_solution_inc})) up to the third order when the
unperturbed star is incompressible. Likewise, we have derived a
recursive formula to find the corrections to the tidal Love
numbers due to an arbitrary density perturbation on a general
density profile (see equation (\ref{eqn:yj_solution})).

Using the density profile obtained by applying the SDEM to
polytropic stars \citep{Kenny_SDEM}, and the pertubative analyses
for the tidal Love numbers and the multipole moments developed in
Sections \ref{sec:general_perturbation} and
\ref{sec:ILoverelations_polytropes} respectively, we have
determined perturbation series about the centres $n=0,1$ as well
as two-point Pad\'{e} approximants for
various physical quantities, 
including $k_2$, $\mathcal{M}_{2} / \mathcal{M}_{0}$, $U_2$,
$k_3$, $\mathcal{M}_{4} / \mathcal{M}_{0}$ and $U_3$. In
particular, as shown in Table~\ref{table:error}, the two-point
Pad\'{e} approximants for these quantities are very accurate.
Throughout the entire physical range where $0 \le n \le 3$, the
maximum errors for the two-point Pad\'{e} approximants listed in
the table is less than $1$ per cent, except for the case
$U_{3,\text{tp}}$ with a maximum error slightly greater than $2$ 
per cent. As a matter of fact, for stiff polytropic stars with $0 \le n
\le 1$, these two-point Pad\'{e} approximants are nearly exact. To
make Table~\ref{table:error} self-contained, we have also included
the results of the corresponding approximants for the
dimensionless radius in equation (\ref{eqn:appendix_radius})
\citep[see][equation (83)]{Kenny_SDEM}, and the dimensionless mass
in equation (\ref{eqn:appendix_mass}) \citep[see][equation
(86)]{Kenny_SDEM}. One can therefore apply these simple
approximants to accurately evaluate various physical quantities,
including mass, radius, moment of inertia, and tidal Love numbers
(deformabilities), of stable polytropic stars.  For ease of 
reference, a self-contained summary of the accurate two-point approximants 
are given in Appendix \ref{sec:formulae_for_lazy_physicist}. 

Furthermore, as could be seen the accuracy of the values of $U_2$
and $U_3$ shown in Figures \ref{fig:D2_quotient_explanation} and
\ref{fig:D3_quotient_explanation}, and Table~\ref{table:error},
the I-Love relation (or multipolar moment-Love relations) of
polytropic stars is also accurately reproduced by our perturbative
analysis. Consequently, we can study these relations analytically.
Both the analytical and numerical results show that the variation
of the I-Love relation (i.e., $U_2$) is within $0.4$ per cent near the
incompressible limit where $n\in[0,1]$, growing sharply to $6$ per cent
in softer EOS when $n=2$, and reaching $25$ per cent in the stability
limit when $n=3$. Moreover, the universality of the I-Love
relation around the incompressible limit is found to be
attributable to the existence of the universal stationary point at
the incompressible limit $n=0$, and a secondary stationary point
around $n \approx 0.4444$. The existence of such a secondary
stationary point is in good agreement with the Taylor expansion of
$U_2$ (see equation \eqref{eqn:U2n0}). The octupolar moment-Love
relation of polytropes ($l=3$) is analysed in the same fashion,
and it is observed to vary more significantly for the absence of a
secondary stationary point. This signifies that the universality
of the multipolar moment-Love relations would degrade as the
angular momentum index $l$ increases.

\begin{table}
\centering \caption{\label{table:error} The table summarises the
maximum percentage errors near the incompressible limit for
$n\in[0,1]$ and over the range of stable polytropes for
$n\in[0,3]$, for the two-point Pad\'{e} approximant of the tidal
Love numbers of degree $l=2$ and $l=3$ in equation
(\ref{eqn:klPade}), the moment fraction for $l=2$ and $l=3$ in
equation (\ref{eqn:moment_ratio_tp}), the I-Love relation
$U_{2,\text{tp}} = k_{2,\text{tp}} [\mathcal{M}_{2} /
\mathcal{M}_0]_{\text{tp}}^{-5/2}$, the octupole moment-Love
relation $U_{3} = k_{3,\text{tp}} [\mathcal{M}_{4} /
\mathcal{M}_0]_{\text{tp}}^{-7/4}$, the dimensionless radius in
equation (\ref{eqn:appendix_radius}) \citep[see][equation
(83)]{Kenny_SDEM}, and the dimensionless mass in equation
(\ref{eqn:appendix_mass}) \citep[see][equation (86)]{Kenny_SDEM}.
}
 \begin{tabular}{|l |l | l | l |}
    \hline
    Quantity                            &   Two-point approximant                           & $0 \leq n \leq 1$         & $0 \leq n \leq 3$     \\
    \hline
    $k_{2}$                             & $k_{2,\text{tp}}$                                 & $2.5 \times 10^{-5} \%$   & $0.39 \%$             \\
    $\mathcal{M}_{2} / \mathcal{M}_{0}$ & $[\mathcal{M}_{2} / \mathcal{M}_0]_{\text{tp}}$   & $6.1 \times 10^{-5} \%$   & $0.21 \%$             \\
    $U_{2}$                             & $U_{2,\text{tp}}$                                 & $1.3 \times 10^{-4} \%$   & $0.93 \%$             \\
    \hline
    $k_{3}$                             & $k_{3,\text{tp}}$                                 & $7.5 \times 10^{-5} \%$   & $0.93 \%$             \\
    $\mathcal{M}_{4} / \mathcal{M}_{0}$ & $[\mathcal{M}_{4} / \mathcal{M}_0]_{\text{tp}}$   & $1.6 \times 10^{-4} \%$   & $0.65 \%$             \\
    $U_{3}$                             & $U_{3,\text{tp}}$                                 & $3.5 \times 10^{-4} \%$   & $2.1 \%$              \\
    \hline
    $R/a$                               & $\xi_{\text{g}}$                                  & $8.1 \times 10^{-7} \%$   & $0.022 \%$            \\
    $M/(\rho_{c} a^3)$                  & $m_{\text{g}}$                                    & $8.5 \times 10^{-5} \%$   & $0.091 \%$            \\
    \hline
 \end{tabular}
\end{table}

The form of the multipolar moment-Love relation
\eqref{eqn:Uldefinition} is also worthy of discussion. As $U_l$ is
universal, it is expected to be independent of the spatial size
and the mass of a star. It is obvious that $U_l$ is dimensionless
and equation \eqref{eqn:Uldefinition} is dimensionally correct.
However, one can similarly construct other dimensionally correct
formulae, such as
\begin{equation} \label{eqn:Vldefinition}
V_{l} \equiv  k_{l}
\left(\frac{\mathcal{M}_{2l-2}}{\mathcal{M}_{0}}\right)^{-\chi},
\end{equation}
where $V_{l}$ and $\chi$ are two dimensionless constants. How can
one determine the value of  $\chi $ such that the above formula is
EOS-insensitive?

The answer  to this question lies in the physical properties  of
$k_l$ and ${\cal M}_k$, which are defined by equations
\eqref{eqn:kldef} and
 \eqref{eqn:M_k}, respectively. In order to make the quantity $V_{l}$
 EOS-insensitive, the RHS of  \eqref{eqn:Vldefinition} should remain unchanged if a  mass shell of zero (or low) density
 is artificially added on
 a specific star. However, the radius of such a composite star increases from $R$ to $R +
 \Delta R$, where  $\Delta R$ is the thickness of the mass shell. On the other hand, such a  massless shell cannot  change
 the values of $q_l$ and $I_k$ appearing in equations \eqref{eqn:kldef} and
 \eqref{eqn:M_k}, respectively. Physically speaking, both $q_l$ and $I_k$ are global dynamical variables
 of the star and should not depend on an artificial increment in the stellar radius due to a variation in the
 extremely low density portion of the relevant EOS.
 As $k_l \propto q_l/R^{2l+1}$ and ${\cal M}_k \propto
 I_k/R^{k+3}$, it is obvious that the value of $V_{l}$ will be
 affected by this zero-density mass shell unless $\chi =
 (2l+1)/(2l-2)$. Thus,   the correct form of the I-Love relation (or the multipolar moment-Love
 relation),  \eqref{eqn:Uldefinition}, can be obtained.

 On the other hand, one could also suggest another possible form
 of the universal relation:
\begin{equation} \label{eqn:Vldefinition-2}
V_{l,m} \equiv  k_{l}
\left(\frac{\mathcal{M}_{2m-2}}{\mathcal{M}_{0}}\right)^{-(2l+1)/(2m-2)},
\end{equation}
where $m$ is a positive integer other than $l$. The above formula
seems to be correct because (i) it is dimensionally correct, and
(ii) remains invariant upon the introduction of a massless shell
as mentioned above. In fact, this form has been considered by
\citet{Samson_fmode_ILoveQ} and was coined as the off-diagonal
relation to signify that $l \neq m$. However, it follows directly
from the derivation developed in
Section~\ref{sec:ILoverelations_general} that the so defined
quantity $V_{l,m}$ is no longer stationary about the
incompressible limit where $n=0$. As shown in equations
\eqref{eqn:klv} and \eqref{eqn:Mlv}, in the limit $\varepsilon=0$,
$d \ln k_l/d \varepsilon$ consists of two contributions, namely,
$\mu_0(R)$ and $\mu_{2l-2}(R)$, while $d \ln {\cal M}_{2m-2} /d
\varepsilon$ is proportional to $\mu_{2m-2}(R)$. It is then
obvious that, in general,  $ d V_{l,m}/d \varepsilon$ does not
vanish unless $m=l$. Therefore, the relation defined in
\eqref{eqn:Vldefinition-2} is not stationary about the
incompressible limit. We conclude that such a relation would not
hold universally for compact stars with good precision. As a
matter of fact, our conclusion is consistent with the numerical
result obtained by \citet{Samson_fmode_ILoveQ}.

The scope of the present paper  is confined to Newtonian gravity. 
However, the approximants of the tidal Love numbers and mass moments 
of polytropic stars are accurate over a wide range of polytropic indices. 
They are useful to astrophysicists in their own right. 
In addition, they also provide a physically transparent picture 
for the I-Love-Q relations, at least in the Newtonian regime. 
We expect that similar investigation could also be performed 
in Einstein gravity in due course. 

\section*{Acknowledgements}
We sincerely thank
L. M. Lin, S. Y. Lau, H. F. Shek, K. K. Lee, 
Simon K. S. Chu, Nicole S. M. Luk, K. T. Tai and P. T. Fong 
at the Physics Department, the Chinese University of Hong Kong, as well as
T. K. Chan at the Department of Physics, the University of California at San Diego,
for fruitful discussion and valuable suggestions.  
We particularly thank C. F. Lo for his ingenious suggestion of generalising the 
analysis of the tidal Love numbers, and subsequently the I-Love-Q relation to 
an EOS-independent form, through starting from an arbitrary variation of the density profile instead of the EOS.    
We also thank Tjonnie G. F. Li, Peter T. H. Pang, Otto A. Hannuksela, and Rico K. L. Lo, 
members of LIGO Scientific Collaboration group, at the Physics Department, the Chinese University of Hong Kong,
for the discussion about the LIGO experiment and gravitational wave physics.




\bibliographystyle{mnras}
\bibliography{Love_numbers_references} 




\appendix

\section{Perturbative solution to the scaled polytrope function in SDEM} \label{sec:SDEM_formulae_Theta_only}
The perturbative solution to the scaled polytrope function at the perturbation centres $n=0$ and $n=1$ are summarized below.  For details, readers may refer to a recent article proposing the SDEM in LEE \citep{Kenny_SDEM}.
\begin{equation}
\Theta_{0}^{(0)} (z) = 1 - \frac{z^2}{\pi^2},
\end{equation}
\begin{align}
\nonumber \Theta_{0}^{(1)}(z) = & -4 + \frac{4 z^2}{\pi^2} \left(1 - \ln 2 \right)
+\left(3-\frac{2 \pi }{z}-\frac{z^2}{\pi ^2}\right) \ln \left(1-\frac{z}{\pi}\right)\\
& +\left(3+\frac{2 \pi }{z}-\frac{z^2}{\pi ^2}\right) \ln
\left(1+\frac{z}{\pi }\right),
\end{align}
\begin{align}
\Theta_{0}^{(2)}(z) = & 40+\frac{7 \pi ^2}{3}+8 \ln 2-14 \ln ^2 2 \nonumber \\
& + \left[\frac{7 \pi^2}{3}- 40 +32 \ln 2 -8 \ln ^2 2 \right] \frac{z^2}{\pi^2} \nonumber \\
& + \left[-23+\frac{20 \pi }{z}+\frac{3 z^2}{\pi ^2}+ \big(14 -\frac{10 \pi}{z}-\frac{4 z^2}{\pi ^2}\big)\ln 2 \right]\nonumber \\
& \quad \times \ln \left(1-\frac{z}{\pi}\right) + \left(\frac{3}{2}-\frac{\pi }{z}-\frac{z^2}{2 \pi ^2}\right) \ln ^2\left(1-\frac{z}{\pi }\right) \nonumber \\
& + \left[-23-\frac{20 \pi }{z}+\frac{3 z^2}{\pi ^2}+\big( 14+\frac{10 \pi}{z}-\frac{4 z^2}{\pi ^2}\big) \ln 2 \right] \nonumber \\
& \quad \times \ln \left(1+\frac{z}{\pi}\right) + \left(\frac{3}{2}+\frac{\pi }{z}-\frac{z^2}{2 \pi^2}\right) \ln ^2\left(1+\frac{z}{\pi }\right) \nonumber \\
& + \left(1-\frac{z^2}{\pi ^2}\right) \ln \left(1-\frac{z}{\pi }\right) \ln \left(1+\frac{z}{\pi }\right)\nonumber \\
& + \left(\frac{14 \pi }{z}-14\right) \text{Li}_2\left(\frac{\pi -z}{2 \pi }\right)\nonumber \\
& + \left(-\frac{14 \pi }{z}-14\right) \text{Li}_2\left(\frac{\pi
+z}{2 \pi}\right),
\end{align}
\begin{equation}
\Theta_{1}^{(0)}(z) = \frac{\sin z}{z},
\end{equation}
\begin{align}
\Theta_{1}^{(1)}(z) & = \frac{\sin z}{z} \bigg[ 1-\frac{\ln(2\pi)}{2}-\frac{\text{Si}(2 \pi )}
{4 \pi } - \frac{\text{Cin}(2 z)}{4} \nonumber \\
& +\frac{\ln z}{2}-\frac{1}{2} \ln(\sin z) \bigg] + \frac{\cos z}{z} \bigg[ \frac{1}{2} z \ln (2 \pi) \nonumber \\
& -\frac{1}{2} z \ln z+\frac{z \text{Si}(2 \pi )}{4 \pi
}-\frac{\text{Si}(2 z)}{4} + \frac{1}{2} \int_{0}^{z} \ln \sin t
dt \bigg].
\end{align}
Here $\text{Li}_2(z)$, called the dilogarithm (or polylogarithm of order $2$), is defined by \citep[see,e.g.,][]{Math_handbook,Lewin_polylog}:
\begin{equation}
\text{Li}_2 (z) = \int_{0}^{z} \frac{ - \ln(1-t) }{t} dt,
\end{equation}
 $\text{Si}(x)$ and $\text{Cin}(x)$ are integrals of sine and
cosine defined respectively by \citep[see,
e.g.,][]{Math_handbook}:
\begin{equation}\label{eq:Si}
\text{Si}(x) = \int_{0}^{x} \frac{\sin t}{t} dt,
\end{equation}
\begin{equation}
\text{Cin}(x) = \int_{0}^{x} \frac{1 - \cos t}{t} dt.
\end{equation}

\section{Asymptotic behaviour of tidal Love number and moments of soft polytropes}\label{sec:n5}
Although we are primarily interested in stable polytropic stars
with $n \in [0,3]$ in the present paper, we find that the
asymptotic behaviour of tidal Love numbers and moments around the
case $n =5$, whose LEE is also exactly solvable, is noteworthy. In
both the two-point approximants \eqref{eqn:klPade} and
\eqref{eqn:moment_ratio_tp}, we have included an empirical factor
$(n-5)^3$ to improve their accuracy. Such a factor can be argued
from the above-mentioned asymptotic behaviour and extrapolation.
In the following we briefly sketch our argument.

The LEE for polytropes with $n =5$ can be solved exactly with
$\theta(x)= (1+x^2/3)^{1/2}$. Hence, for soft polytropes with $n
\approx 5$ and $x \gg 1$, we have the following approximate
dependence: (i) $\theta(x) \propto 1/x$, (ii) $\rho \propto
1/x^5$, and (iii) $P \propto 1/x^6$. As a result, the induced mass
density in the Poisson's equation \eqref{eqn:H_definition} is
proportional to $1/x^4$. On the other hand, the potential at a
large distance is dominated by the external tidal field, which go
as $x^l$. Using standard Green's function method and the multipole
expansion of $1/|\bf{r}-\bf{r}'|$ \citep[see, e.g.,][]{JacksonEM},
\begin{equation}\label{Coul}
\frac{1}{|\bf{r}-\bf{r}'|} \equiv
\sum_{l=0}^{\infty}\frac{r_<^l}{r_>^{l+1}}P_l(\cos \gamma),
\end{equation}
where $\bf{r}$ and $\bf{r}'$ are two arbitrary 3-dimensional
vectors, $\gamma$ is the angle formed between these two vectors,
$r_> = {\rm max} (|\bf{r}|,|\bf{r}'|)$ and $r_< = {\rm min}
(|\bf{r}|,|\bf{r}'|)$,   it is straightforward to show that the
induced multipole potential outside the star is given by a term
proportional to  $R^{2l-1} /r^{l+1}$, implying that the induced
multipole moment $q_l$ is proportional to $R^{2l-1}$. As the tidal
Love number $k_l$ is given by $q_l /(2 R^{2l+1})$, we see that
$k_l \propto 1/R^2$.

Given that $R \propto 1/(n-5)$  as $n \rightarrow 5$
\citep{Buchdahl_n5}, we can show that $k_l \propto (n-5)^2$ under
the same limit. We have confirmed this analytic result by
comparing it with  our numerical results. However, we are
interested in stable stars with $n \in [0,3]$. We found
numerically that $k_l$ is approximately proportional $(n-5)^3$ for
$n \approx 3$. Therefore, we have included an empirical factor of
$(n-5)^3$, instead of $(n-5)^2$, in the construction of the
two-point approximant \eqref{eqn:klPade}.

On the other hand, it follows directly from equation
\eqref{eqn:M_k} that the leading dependence of ${\cal M}_2 /{\cal
M}_0$ and ${\cal M}_4 /{\cal M}_0$ are given by $R^{-2} \ln R \sim
R^{-2}$ and $R^{-2}$, respectively, as $R \rightarrow \infty$. To
take care of this behaviour for soft yet stable polytropic stars
with $n \approx 3$, we similarly add an empirical factor of
$(n-5)^3$ in the two-point approximant
\eqref{eqn:moment_ratio_tp}. Once again we see from the results
discussed in Section~\ref{sec:ILoverelations_polytropes} that the
introduction this factor can indeed significantly improves the
accuracy of the approximant for soft polytropic stars.

\section{List of Useful Formulae} \label{sec:formulae_for_lazy_physicist}

For ease of reference, in the following we quote the approximate
expressions of the radius, the tidal Love numbers of degree $l=2$
and $l=3$, the mass, the moment of inertia and the octupole moment
of polytropes, as a function of the polytropic index $n$.

\subsection{Radius}
For a polytrope of central density $\rho_c$ under the polytropic
equation of state $P(r)=K[\rho(r)]^{1+1/n}$, its radius is equal
to \citep{Kenny_SDEM}:
\begin{equation}
R = \sqrt{\frac{K (n+1)}{4 \pi G}} \rho_{c}^{(1-n)/(2n)} \hat{\xi}(n) = a \hat{\xi}(n), \\
\end{equation}
where $\hat{\xi}$ is the first zero of the the normalised
polytrope function, approximately given by the following
approximant:
\begin{equation}
\label{eqn:appendix_radius} \hat{\xi}_{\text{g}}(n) = \frac{ \pi
}{1 + h_{0} n^{8}(n-1)^{8}} \bigg[ \frac{h_{1} + h_{2} n + h_{3}
n^2 + h_{4} n^3}{ \sqrt{5-n} \big(1 + h_{5} n +  h_{6} n^2 \big)}
\bigg]^{(n-1)/2},
\end{equation}
with the constants $h_{i}$ for $i=0,1,2,\ldots,6$ determined by
the SDEM as follows:
\begin{equation*}
\begin{split}
& h_{0} \approx 1.599664440540 \times 10^{-17}, \\
& h_{1} \approx 3.67818439198, \\
& h_{2} \approx  -0.1212783778520, \\
& h_{3} \approx -0.082089876683, \\
& h_{4} \approx 0.003032776676846, \\
& h_{5} \approx 0.0085827378725, \\
& h_{6} \approx -0.01884581518309.
\end{split}
\end{equation*}
The error of $\hat{\xi}_{\text{g}}(n)$ is within $8.1 \times
10^{-7}$ per cent near the incompressible limit for $n\in[0,1]$,
and $0.022$ per cent throughout the physical range of stable
polytropes for $n\in[0,3]$ \citep[see][Figure 6]{Kenny_SDEM}.

\subsection{Tidal deformabilities and tidal Love numbers}
The tidal Love numbers $k_{l}$ and the tidal deformabilities
$\lambda_{l}$ are related by:
\begin{equation}
\lambda_{l} = \frac{2}{(2l-1)!!} R^{2l+1} k_{l}.
\end{equation}
For the polytropic equation of state, $k_{2}$ and $k_{3}$ are
approximated by:
\begin{align}
\label{eqn:klPade_appendix} k_{l,\text{tp}}(n) = (5-n)^3
\frac{a_{1} + n a_{2} + n^2 a_{3} + n^3 a_{4}}{1000 + n a_{5} +
n^2 a_{6}},
\end{align}
where are constants are listed in Table \ref{table:appendix_1}.
The percentage errors are, respectively, within $2.5 \times
10^{-5} $ per cent and $7.5 \times 10^{-5} $ per cent near the
incompressible limit for $n\in[0,1]$, and $0.39$ per cent and
$0.93$ per cent for stable polytropes over the range $n\in[0,3]$.
The details could be could in Figures
\ref{fig:k2_convergence_accuracy} and
\ref{fig:k3_convergence_accuracy}.
\begin{table}
\caption{\label{table:appendix_1} The numerical values of the
constants $a_{1}, a_{2}, \ldots, a_{6}$ appearing in
\eqref{eqn:klPade_appendix}.} \centering
 \begin{tabular}{|l |l | l |}
    \hline
            & $l=2$             & $l=3$ \\
    \hline
    $a_{1}$ & $6$                   & $3$                   \\
    $a_{2}$ & $0.0701474990944$     & $-0.0540432029285$   \\
    $a_{3}$ & $-0.292565896945$     & $-0.248516990271$     \\
    $a_{4}$ & $0.0269208110373$     & $0.0346247187561$      \\
    $a_{5}$ & $411.691249849$       & $581.985599024$       \\
    $a_{6}$ & $17.6102741235$       & $60.5271512307$       \\
    \hline
 \end{tabular}
\end{table}

\subsection{Mass}
The physical mass $M$ of the same polytrope is given by
\citep{Kenny_SDEM}:
\begin{equation}
M = \int_{0}^{R} \rho(r) 4 \pi r^2 dr = \left[\frac{(n+1)K}{4\pi
G}\right]^{3/2} \rho_{c}^{(3-n)/(2n)} m(n),
\end{equation}
where $m$ is the dimensionless mass approximately equal to
$m_{\text{g}}(n)$:
\begin{align}
m_{\text{g}}(n) = & j_{0} (5-n)^{(15-3n)/4} n^8 (n-1)^8 \nonumber \\
& + 4 \pi \bigg[ \frac{h_{1} + h_{2} n + h_{3} n^2 + h_{4} n^3}{ \sqrt{5-n} \big(1 + h_{5} n +  h_{6} n^2 \big)} \bigg]^{3(n-1)/2} \nonumber \\
\label{eqn:appendix_mass} & \times (5-n)^3 \frac{j_{1} + n j_{2} +
n^2 j_{3} + n^3 j_{4}}{1 + n j_{5} + n^2 j_{6}},
\end{align}
where the constants $h_{i}$ for $i=1,2,...,6$ are given above and
the constants $j_{i}$ for $i=0,1,2,...,6$ are determined by the
SDEM:
\begin{equation*}
\begin{split}
& j_{0} \approx -3.42086751650\times 10^{-10}, \\
& j_{1} \approx 0.0826834044808, \\
& j_{2} \approx  0.0570923774428, \\
& j_{3} \approx -0.00213715241113, \\
& j_{4} \approx -0.000863277094516, \\
& j_{5} \approx 1.37086609691, \\
& j_{6} \approx 0.415498502167.
\end{split}
\end{equation*}
The error is within $8.5 \times 10^{-5}$ per cent for $n \in
[0,1]$, and $0.091$ per cent throughout the range of stable
polytropes for $n\in[0,3]$ \citep[see][Figure 7]{Kenny_SDEM}.

\subsection{Moment of inertia and octupole moment}
Define the $k$th moment of the density profile by:
\begin{equation}
I_{k} = \int_{0}^{R} \rho(r) r^{k+2} dr,
\end{equation}
where the $r^2$ factor accounts for the spherical volume element.
Note that the total mass $M = 4 \pi I_{0}$, and the moment of
inertia $I = (8\pi/3) I_{2}$. For $l=2$ and $l=3$, accurate
two-point Pad\'{e} approximant is determined by the SDEM:
\begin{equation}
\label{eqn:appendix_moment_fraction} \bigg[\frac{I_{2l-2}}{I_{0}
R^{2l-2}}\bigg]_{\text{tp}} = (5-n)^3 \frac{b_{1} + n b_{2} + n^2
b_{3}  + n^3 b_{4} }{1000 + n b_{5} + n^2 b_{6}},
\end{equation}
where the constants $b_{i}$ for $i=1,2,\ldots, 6$ are listed in
Table \ref{table:appendix_2}.
\begin{table}
\centering \caption{\label{table:appendix_2} The numerical values
of the constants $b_{1}, b_{2}, \ldots, b_{6}$ appearing in
\eqref{eqn:moment_ratio_tp}.}
 \begin{tabular}{|l |l | l |}
    \hline
            & $l=2$                 & $l=3$ \\
    \hline
    $b_{1}$ & $24/5$                & $24/7$                 \\
    $b_{2}$ & $3.452950061777$      & $1.605541075759$      \\
    $b_{3}$ & $0.08417922198458$    & $0.5058524906146$      \\
    $b_{4}$ & $0.01816712663676$    & $0.02106526231457$      \\
    $b_{5}$ & $519.3645962035$      & $553.9970994773$     \\
    $b_{6}$ & $-155.4882028232$     & $-137.1355972892$     \\
    \hline
 \end{tabular}
\end{table}
The percentage errors of the moment fractions for $l=2$ and $l=3$
are, respectively, within $6.1 \times 10^{-5}$ per cent and $1.6
\times 10^{-4}$ per cent near the incompressible limit for
$n\in[0,1]$, and $0.21$ per cent and $0.65$ per cent for stable
polytropes over the range $n\in[0,3]$. The details could be could
in Figures \ref{fig:M2oM0} and \ref{fig:M4oM0}.  Together with the
aforementioned approximation to the radius and the mass, the
expressions yield accurate approximation to the moment of inertia
and the octupole moment.


\bsp    
\label{lastpage}
\end{document}